\pdfoutput=1
\documentclass[11pt,twoside,a4paper,cmspaper,final,collab]{cms-tdr}
\def\svnVersion{8136a0d}\def\svnDate{2022/02/28}\def\cmsCernNoTag{CERN-EP-2021-241}\def\cmsCernDate{\today}\def\cmsMessage{Published in the Journal of High Energy Physics as \href{http://dx.doi.org/10.1007/JHEP02(2022)169}{\doi{10.1007/JHEP02(2022)169}.}}
\begin{document}\cmsNoteHeader{TOP-19-002}

\newcommand{\ttbb}{\ensuremath{\ttbar{}\bbbar}\xspace}
\newcommand{\ttcc}{\ensuremath{\ttbar{}\ccbar}\xspace}
\newcommand{\ttlf}{\ensuremath{\ttbar{}\cmsSymbolFace{LF}}\xspace}
\newcommand{\tW}{\ensuremath{\PQt{}\PW}\xspace}
\newcommand{\mtop}{\ensuremath{m_{\PQt}}\xspace}
\newcommand{\hdamp}{\ensuremath{h_{\text{damp}}}\xspace}
\newcommand{\Hq}{\ensuremath{\PH\PQq}\xspace}
\newcommand{\Hqt}{\ensuremath{\PH\PQq\PQt}\xspace}
\newcommand{\Hut}{\ensuremath{\PH\PQu\PQt}\xspace}
\newcommand{\Hct}{\ensuremath{\PH\PQc\PQt}\xspace}
\newcommand{\ttoHu}{\ensuremath{\PQt \to \PH\PQu}\xspace}
\newcommand{\ttoHc}{\ensuremath{\PQt \to \PH\PQc}\xspace}
\newcommand{\ttoHq}{\ensuremath{\PQt \to \PH\PQq}\xspace}
\newcommand{\Htobb}{\ensuremath{\PH \to \bbbar}\xspace}
\newcommand{\kappaHqt}{\ensuremath{\kappa_{\Hqt}}\xspace}
\newcommand{\kappaHut}{\ensuremath{\kappa_{\Hut}}\xspace}
\newcommand{\kappaHct}{\ensuremath{\kappa_{\Hct}}\xspace}
\newcommand{\cmsTable}[1]{\resizebox{\textwidth}{!}{#1}}

\cmsNoteHeader{TOP-19-002}
\title{Search for flavor-changing neutral current interactions of the top quark and the Higgs boson decaying to a bottom quark-antiquark pair at \texorpdfstring{$\sqrt{s}=13\TeV$}{sqrt(s) = 13 TeV}}

\date{\today}

\abstract{
A search for flavor-changing neutral current interactions of the top quark (\PQt) and the Higgs boson (\PH) is presented. The search is based on a data sample corresponding to an integrated luminosity of 137\fbinv recorded by the CMS experiment at the LHC in proton-proton collisions at $\sqrt{s}=13\TeV$. Events containing exactly one lepton (muon or electron) and at least three jets, among which at least two are identified as originating from the hadronization of a bottom quark, are analyzed. A set of deep neural networks is used for kinematic event reconstruction, while boosted decision trees distinguish the signal from the background events. No significant excess over the background predictions is observed, and upper limits on the signal production cross sections are extracted. These limits are interpreted in terms of top quark decay branching fractions ($\mathcal{B}$) to the Higgs boson and an up (\PQu) or a charm quark (\PQc). Assuming one nonvanishing extra coupling at a time, the observed (expected) upper limits at 95\% confidence level are $\mathcal{B}(\ttoHu) < 0.079\,(0.11)\%$ and $\mathcal{B}(\ttoHc) < 0.094\,(0.086)\%$.
}

\hypersetup{
pdfauthor={CMS Collaboration},
pdftitle={Search for flavor-changing neutral current interactions of the top quark and the Higgs boson decaying to a bottom quark-antiquark pair at sqrt(s)=13 TeV},
pdfsubject={CMS},
pdfkeywords={CMS, top quark, FCNC, Higgs}}

\maketitle

\section{Introduction}
\label{sec:introduction}

The discovery of a new scalar boson with properties compatible with the standard model (SM) Higgs boson (\PH)~\cite{higgs2, higgs1, higgs3, higgs4} opened a new window to shed light on beyond-the-SM phenomena at the CERN LHC.
The top quark (\PQt) is the most massive known fermion and its measured Yukawa coupling is of order unity~\cite{CMS:2020djy}.
The large mass of the top quark (\mtop) suggests that it may play an important role in electroweak symmetry breaking~\cite{Dobrescu:1997nm,Chivukula:1998wd,Delepine:1995qs}.
Since the Yukawa coupling of the Higgs boson to fermions is proportional to the mass of the fermion, it is of particular importance to study the coupling between the top quark and Higgs boson.

Flavor-changing neutral current (FCNC) interactions of the \Hqt vertex (where \PQq represents quarks) are key processes to search for new physics.
The FCNC process is forbidden at tree level in the SM and is highly suppressed in loop corrections by the
Glashow--Iliopoulos--Maiani mechanism~\cite{GIM}, leading to a branching fraction of $\mathcal{O}(10^{-17})$ for \ttoHu and
$\mathcal{O}(10^{-15})$ for \ttoHc~\cite{Aguilar-Saavedra:2004mfd, GIM_BR_2, GIM_BR_3} with the up (\PQu) and charm quark (\PQc) final states, respectively.
On the contrary, many SM extensions, such as the minimal supersymmetric standard model~\cite{FCNC_BSM_3}, supersymmetric models with R-parity violation~\cite{Eilam:2001dh}, and other two-Higgs-doublet models~\cite{Cheng:1987rs,Atwood:1996vj,FCNC_BSM_2}, predict highly enhanced \ttoHq branching fractions~\cite{FCNC_BSM_1, FCNC_BSM_2, FCNC_BSM_3, FCNC_BSM_4, FCNC_BSM_5}.
Values can be as high as $\mathcal{O}(10^{-5})$ for \ttoHu and $\mathcal{O}(10^{-3})$ for \ttoHc~\cite{Aguilar-Saavedra:2004mfd}, which are within the experimental sensitivity of analyses performed with the LHC data at 13\TeV.

Several searches for anomalous couplings between the top quark and the Higgs boson have recently been carried out by the ATLAS~\cite{ATLAS_FCNC_1, ATLAS_FCNC_3, ATLAS_FCNC_4} and CMS~\cite{CMS_FCNC_1,CMS:2021bdg} Collaborations.
The most recent analysis is performed by CMS using a data sample with an integrated luminosity of 137\fbinv in the diphoton decay channel of the Higgs boson, yielding observed limits on the branching fractions of 0.019 and 0.078\% for \ttoHu and \ttoHc, respectively~\cite{CMS:2021bdg}.

\begin{figure}[ht!]
    \centering
    \includegraphics[width=0.48\textwidth]{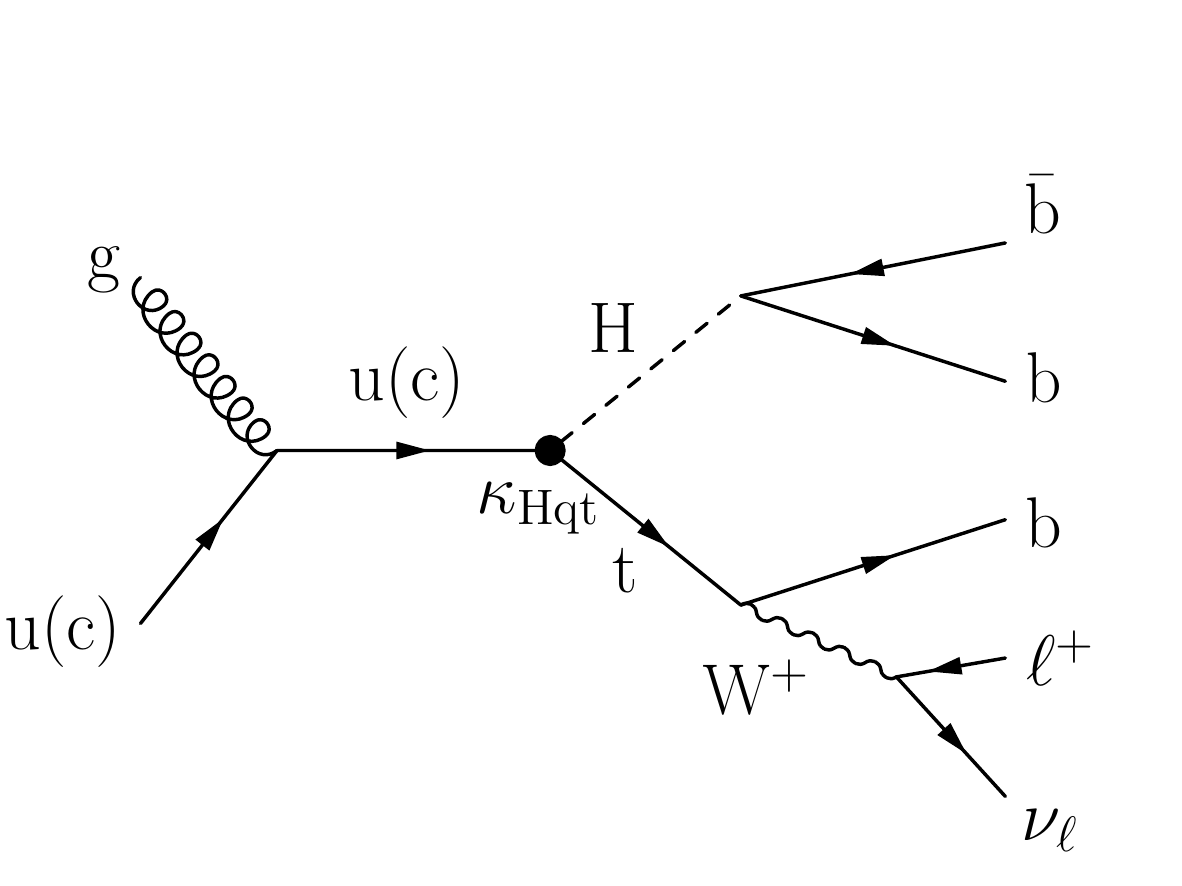}
    \includegraphics[width=0.48\textwidth]{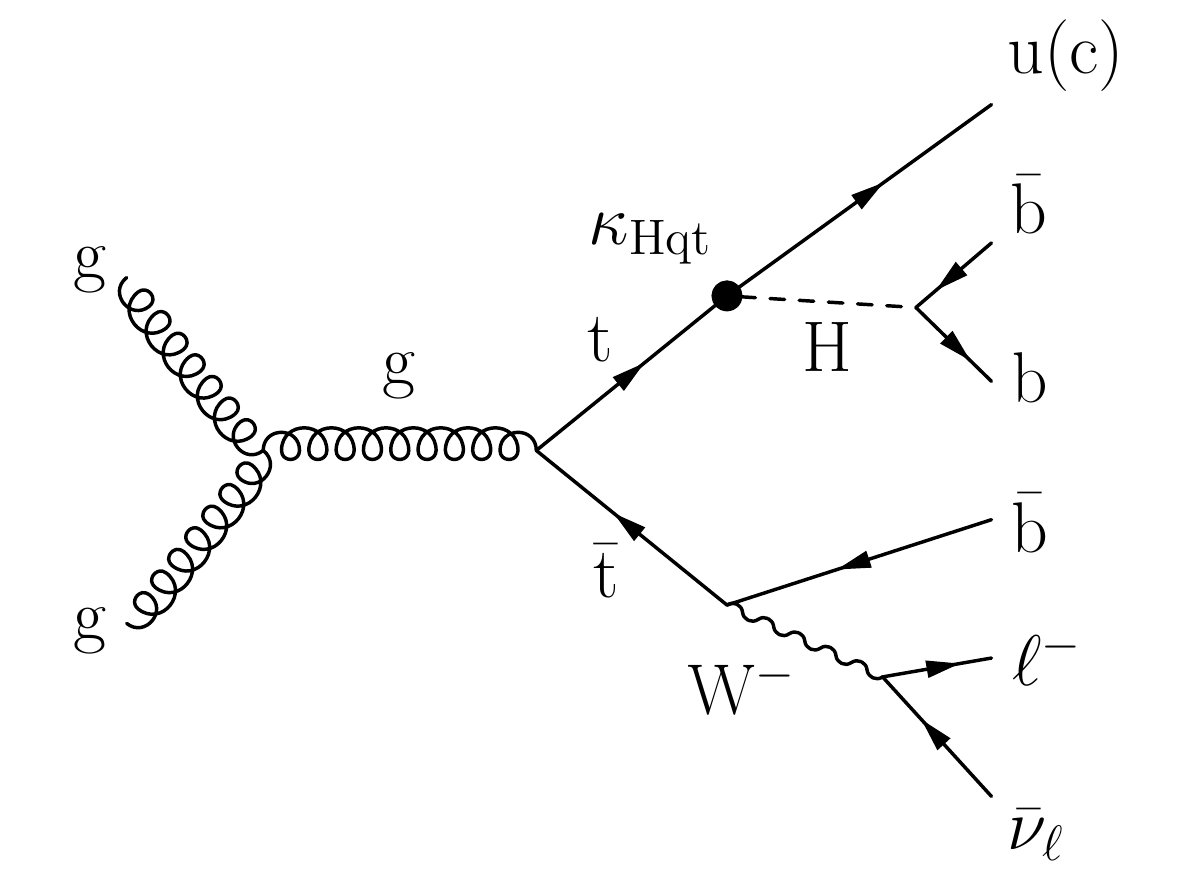}
    \caption{Illustrative Feynman diagrams for \Hqt FCNC interactions: the production of a single top quark with the Higgs boson (left), and FCNC decay of the top quark in \ttbar events (right), where $\PQq = \PQu$ or \PQc.}
  \label{fig:feyn}
\end{figure}

In this paper, we present an extension of the CMS search for \Hqt FCNC interactions in the \Htobb final state~\cite{CMS_FCNC_1}, exploiting a 13\TeV data set corresponding to an integrated luminosity of 137\fbinv.
Proton-proton ($\Pp\Pp$) collision data collected by CMS during 2017 and 2018 (101\fbinv) are analyzed separately and combined with the earlier results in the \Htobb channel based on the 2016 data-taking period.
Events are selected in the final state with exactly one lepton ($\ell=\mu$ or \Pe) and at least three jets, of which at least two must be identified as coming from the hadronization of a bottom quark (\PQb jets).
Two signal production modes arising from the anomalous \Hqt interaction are considered: the associated production of a single top quark with the Higgs boson from an u or c quark (ST production mode) and the \ttoHq FCNC decays of a top quark in top quark-antiquark pair (\ttbar) events (TT production mode), as presented in Fig.~\ref{fig:feyn}.
In addition to the increased integrated luminosity, the analysis utilizes advanced techniques including a deep neural network (DNN) and boosted decision trees (BDTs) to improve the performance of event reconstruction and signal extraction, and in the \Htobb final state it places the most stringent limits to date on the \ttoHu and \ttoHc branching fractions.
Tabulated results are provided in the HEPData record for this search~\cite{hepdata}.

The paper is organized as follows. We begin with a brief overview of the CMS detector in Section~\ref{sec:detector}, followed by a description of the signal and background modeling in Section~\ref{sec:samples}.
The event selection and signal extraction using multivariate analysis techniques are detailed in Sections~\ref{sec:objects} and~\ref{sec:recomva}, respectively.
Numerous systematic uncertainties contribute to the analysis---these are explained in Section~\ref{sec:systematics}.
Finally, the results of the search are presented in Section~\ref{sec:results} and we conclude with a summary in Section~\ref{sec:summary}.

\section{The CMS detector}
\label{sec:detector}

The central feature of the CMS apparatus is a superconducting solenoid of 6\unit{m} internal diameter, providing a magnetic field of 3.8\unit{T}.
Within the solenoid volume, there are a silicon pixel and strip tracker, a lead tungstate crystal electromagnetic calorimeter (ECAL), and a brass and scintillator hadron calorimeter (HCAL), each composed of a barrel and two endcap sections.
Forward calorimeters extend the pseudorapidity ($\eta$) coverage provided by the barrel and endcap detectors. Muons are detected in gas-ionization chambers embedded in the steel flux-return yoke outside the solenoid.
Events of interest are selected using a two-tiered trigger system. 
The first level is composed of custom hardware processors and selects events at a rate of around 100\unit{kHz}~\cite{Sirunyan:2020zal}. The second level, known as the high-level trigger, is implemented in software running on a processor farm and reduces the event rate to around 1\unit{kHz} before data storage~\cite{CMS:2016ngn}.
A more detailed description of the CMS detector, together with a definition of the coordinate system used and the relevant kinematic variables, can be found in Ref.~\cite{Chatrchyan:2008zzk}.

\section{Signal and background modeling}
\label{sec:samples}

The signal samples are generated at leading order (LO) in perturbative quantum chromodynamics (QCD) with \MGvATNLO 2.6.0 (TT production mode) or 2.4.2 (ST production mode)~\cite{Alwall:2014hca}.
The Lagrangian with a FCNC coupling (\kappaHqt)
\begin{equation}
	\mathcal{L} = \sum_{\PQq=\PQu,\PQc}\frac{g}{\sqrt{2}}\,\PAQt\,\kappaHqt(f^\mathrm{L}_{\Hq}P_\mathrm{L} + f^\mathrm{R}_{\Hq}P_\mathrm{R})\PQq\PH+ \text{h.c}. \label{lagrangian}
\end{equation}
is implemented in the F\textsc{eyn}R\textsc{ules} package~\cite{Alloul:2013bka}, and the universal F\textsc{eyn}R\textsc{ules} output format~\cite{Degrande:2011ua} is used to generate the signal model.
The complex chiral parameters $f^\mathrm{L}_{\Hq}$ and $f^\mathrm{R}_{\Hq}$ have negligible impact on the kinematics, and are fixed to 0 and 1, respectively.
Up to two additional partons are generated at the matrix-element level for the TT production mode, while none are considered for the ST mode as they would induce an overlap between the two processes.
The MLM~\cite{Alwall:2007fs} matching prescription is used to interface the hard-process simulation with the parton-shower modeling.
The ST and TT processes are generated for two scenarios (\Hut or \Hct), assuming the presence of only one nonvanishing FCNC coupling at a time (\kappaHut or \kappaHct).
The TT signal production cross section predicted to 46.5\unit{pb} for each coupling, obtained by multiplying the SM \ttbar production cross section by the corresponding branching fraction of the \ttoHq decay, computed by \MGvATNLO.
The predicted \ttbar production cross section is 832\unit{pb}, as calculated with the \textsc{top++} program at next-to-next-to-LO (NNLO) in perturbative QCD including soft-gluon resummation to next-to-next-to-leading-logarithmic order~\cite{Czakon:2011xx}.
The cross section of the subdominant ST production mode is 13.8 (1.9)\unit{pb} when setting the \Hut (\Hct) coupling to unity, computed by \MGvATNLO with LO precision.
It is known that the top quark transverse momentum (\pt) spectra in \ttbar production using LO and next-to-LO (NLO) predictions differ from the higher-order theoretical prediction.
The distribution and normalization of the top quark \pt spectrum in the TT signal events are therefore corrected first from LO to NLO by utilizing simulated SM \ttbar samples, described below, and then to the NNLO QCD + NLO electroweak prediction~\cite{Czakon:2017wor,CMS:2021vhb}.

The dominant background process is SM \ttbar production, which is simulated at NLO precision using \POWHEG v2~\cite{Nason:2004rx, Frixione:2007vw, Alioli:2010xd}.
The \ttbar events are categorized based on the flavor of the additional particle-level jets (those not originating from the top quark decays): events where at least one additional \PQb jet is present (\ttbb); events where at least one additional \PQc jet is present that are not categorized as \ttbb events (\ttcc); and other events (\ttlf, where LF stands for light flavor).
Similar to the TT signal samples, the \pt spectrum in the SM \ttbar sample is corrected from the NLO to the higher-order prediction.
In addition, the single-leptonic and dileptonic decays of the SM \ttbb process are corrected for observed underestimations in simulation of about 20\%~\cite{Sirunyan:2020kga}, and the corresponding uncertainties derived from the measurement are assigned.
The subdominant single top quark background is considered in the $t$ channel, $s$ channel, and in association with a \PW boson (\tW).
The $t$-channel production is simulated using \POWHEG v2, while the $s$-channel production is generated with \MGvATNLO and \tW samples are obtained from \POWHEG v1~\cite{Re:2010bp}.
The predicted NLO cross sections are 136\,(81)\unit{pb} for $t$-channel top quark (antiquark) production and 10.3\unit{pb} for $s$-channel production~\cite{Aliev:2010zk, Kant:2014oha}.
The \tW cross section considered is 72\unit{pb}, derived at NNLO~\cite{Kidonakis:2010ux, Kidonakis:2012rm}.
Both the \ttbar and single top quark production cross sections are calculated assuming $\mtop=172.5\GeV$. The uncertainties associated with these predictions are described in Section~\ref{sec:systematics}.
Smaller SM background contributions arise from the following processes: \ttbar production in association with a vector boson (\ttbar{}\PZ and \ttbar{}\PW) or a Higgs boson (\ttbar{}\PH), simulated with \MGvATNLO and \POWHEG v1, respectively; Drell--Yan and \PW boson production, simulated with \MGvATNLO; and diboson (VV) processes generated with \PYTHIA 8.230~\cite{Sjostrand:2014zea}.
In addition, multijet events are simulated with \PYTHIA 8.230 and used in the phase space where the estimated multijet background is nonnegligible. 

The parton distribution function (PDF) set used for the event generation is NNPDF3.1~\cite{Ball:2014uwa}, while the handling of the initial- and final-state radiation, the parton showering, and the hadronization is performed by \PYTHIA 8.230 with the CP5 tune~\cite{Sirunyan:2019dfx}.
Additional inelastic $\Pp\Pp$ interactions within the same or nearby bunch crossings (pileup) are simulated using \PYTHIA and added to the generated process of interest.
The number of such interactions to be added is estimated based on the measured instantaneous luminosity and a total $\Pp\Pp$ inelastic cross section of 69.2\unit{mb}~\cite{CMS:2018mlc}.
The events are corrected to remove differences between the simulated and observed pileup distributions. 
All the outgoing particles are propagated through a detailed simulation of the CMS detector based on the \GEANTfour~\cite{Agostinelli:2002hh} framework, which models the interactions with matter, the energy deposits in sensitive detector layers, and the resulting electronic responses.

\section{Event reconstruction and selection}
\label{sec:objects}

The particle-flow (PF) algorithm~\cite{Sirunyan:2017ulk} aims to reconstruct and identify each individual particle in an event, with an optimized combination of information from the various elements of the CMS detector.
The reconstructed vertex with the largest value of summed physics-object $\pt^2$ is taken to be the primary $\Pp\Pp$ interaction vertex.
The energy of electrons is determined from a combination of the electron momentum at the primary interaction vertex as determined by the tracker, the energy of the corresponding ECAL cluster, and the energy sum of all bremsstrahlung photons spatially compatible with originating from the electron track.
The muon \pt is obtained from the curvature of the corresponding track built using the tracker and muon chambers.
The energy of charged hadrons is determined from a combination of their momentum measured in the tracker and the matching ECAL and HCAL energy deposits calibrated by the response function of the calorimeters to hadronic showers.
The energy of neutral hadrons is obtained from the corresponding corrected ECAL and HCAL energies.
Finally, the missing transverse momentum vector \ptvecmiss is computed as the negative vector \ptvec sum of all the PF candidates in an event, and its magnitude is denoted as \ptmiss~\cite{Sirunyan:2019kia}.

Jets are reconstructed from the PF candidates using the anti-\kt clustering algorithm \cite{Cacciari:2008gp,Cacciari:2011ma} with a distance parameter of 0.4.
The measured momentum in simulation agrees within 5--10\% of the true momentum over the entire \pt spectrum and detector acceptance~\cite{CMS:2011shu}.
To mitigate the pileup effect on the jet momentum, tracks identified to be originating from pileup vertices are discarded and an offset correction is applied to correct for remaining contributions. 
Jet energy corrections are derived from simulation and confirmed with in situ measurements of the energy balance in dijet, multijet, $\PGg$+jet, and leptonic $\PZ$+jet events.
Only jets with $\pt > 30\GeV$ and $\abs{\eta} < 2.4$ are considered and an additional selection is applied to remove spurious jet-like features originating from isolated noise patterns in certain HCAL regions~\cite{CMS:2009nkz}.
Heavy-flavor jets are identified based on the DeepCSV algorithm~\cite{CMS:2017wtu} at a working point characterized by a \PQb-jet identification efficiency of 68\%, a \PQc-jet and a light-quark or gluon jet misidentification rate of 12 and 1\%, respectively.
An event-by-event correction based on the jet flavor content is applied to account for differences between the simulated and observed DeepCSV distributions~\cite{CMS:2017wtu}.
The analysis requires the presence of at least three jets, among which two or more must be identified as \PQb jets.

The analyzed events were recorded using a combination of muon and electron triggers.
Events in the muon channel are selected using a trigger requiring at least one muon with $\pt > 27\,(24)\GeV$ for the 2017 (2018) data-taking period; in the electron channel, triggers used require either at least one electron with $\pt > 35\,(32)\GeV$ or at least one electron with $\pt > 28\GeV$ and $\abs{\eta} < 2.1$, as well as a global scalar sum of jet transverse momenta exceeding $150\GeV$.
The corrections to the muon trigger efficiencies are derived using the "tag-and-probe" technique~\cite{CMS:2011aa} in a data sample enriched with $\PZ \to \ell \ell$ events. To calculate the electron trigger efficiencies in a similar way without bias, orthogonal single-muon data sets are used together with \ttbar events in the single-lepton and dilepton channels.
Muon candidates are selected offline based on the track fit quality, the number of hits associated with the track, and the track impact parameter~\cite{MuonReco}.
For electron candidates, offline requirements on the shower shape and track identification are applied~\cite{CMS:2020uim}.
The analysis only considers leptons with $\pt > 30\GeV$ and $\abs{\eta} < 2.4$.
Additionally, electrons reconstructed in the gap between the ECAL barrel and endcap ($1.44 < \abs{\eta} < 1.57$) are rejected.
To reduce the contribution from leptonic decays of heavy-flavor hadrons, a selection on the isolation parameter $I_{\text{rel}}$ is applied.
This variable is defined as the ratio of the scalar \pt sum of photons, charged and neutral hadrons within a cone of angular distance $\DR=\sqrt{\smash[b]{(\Delta\eta)^2+(\Delta\phi)^2}}=0.4\,(0.3)$ for muons (electrons) to the lepton \pt.
The muon candidates are required to have $I_{\text{rel}} < 0.15$ while for electron candidates the requirement is defined in bins of \pt and $\eta$.
Events with additional leptons, defined using looser identification criteria, are rejected to reduce background contributions containing nonprompt leptons.

\section{Signal extraction}
\label{sec:recomva}

The selected events are divided into five categories based on the jet and \PQb-tagged jet multiplicities in order to optimize the analysis sensitivity.
The number of jets can be exactly three or at least four, and the corresponding categories are denoted as j3 or j4, respectively.
The events with up to four \PQb jets are analyzed, considering misidentification of jets and additional partons in the matrix-element calculation.
Similar to the jet multiplicity, the categories in terms of \PQb jets are indicated as b2, b3, or b4 when the considered event has exactly two, three, or four \PQb jets, respectively.
Events with five or more \PQb jets are not used due to their large uncertainty and small impact on the analysis.
Following these conventions, the jet categories are identified as b2j3, b2j4, b3j3, b3j4, and b4j4.
The analysis follows two steps of multivariate techniques to identify signal events over the SM prediction.
Considering each jet category separately, events are reconstructed using a set of DNNs, one for each relevant signal and background process hypothesis.
For each process hypothesis, the DNN selects the best combination of jets in each event.
Thus, a given event may have more than one jet assignment, depending on the available process hypothesis.
The BDTs are then utilized for signal event discrimination against the SM backgrounds using the reconstructed kinematic variables from the jet combinations from all possible process hypotheses.
The obtained distributions for the simulated samples are then fit to the data using a binned maximum likelihood fit.

To maximize the sensitivity to the signal processes and to discriminate against the SM backgrounds, it is necessary to fully characterize the signal and background kinematic properties and gather discriminative features.
Thus, one has to correctly assign the reconstructed jets to the generator-level jets in the signal or background simulated samples to build the best possible distributions of the kinematic variables.
The event reconstruction is performed under multiple process hypotheses, namely the ST and TT signal hypotheses and the SM \ttbar background hypothesis.
Each hypothesis can be distinguished by the number of jets in the final state and their flavor, without considering jets from additional partons or radiation.
The signal hypotheses commonly contain three \PQb jets from the Higgs boson and the leptonically decaying top quark, and the TT signal process has an additional up or charm quark jet.
For the SM \ttbar hypothesis, there is one \PQb jet from each top quark decay and two additional jets from the hadronic decay of a \PW boson.
Consequently, the ST hypothesis is considered for events with three or more jets, while the TT and \ttbar hypotheses are applied to the events with four or more jets. 

Each reconstruction algorithm is constructed as a binary classifier DNN with several possible combinations of three (ST) or four (TT, \ttbar) jets in the event as inputs.
For each event, the signal combination is defined as the case in which all reconstructed jets in the combination are matched to the generator-level jets within an angular distance of 0.4.
All the other combinations are labeled as background.
To reduce the number of wrong combinations considered, the TT and ST signal hypothesis training for events with two (at least three) \PQb jets is performed only for configurations in which the Higgs boson is associated with at least one (two) \PQb jet(s).
After optimization of the performance, and taking into account agreement between data and simulation, 32 (69) kinematic variables are constructed for the ST (TT and SM \ttbar) hypotheses and used for the training, such as the angular distance, momentum, and $\eta$ of jets, dijets, or trijet combinations as well as \PQb tagging discriminants.

A residual network structure~\cite{DBLP:journals/corr/HeZRS15,Erdmann:2017hra} is chosen to boost training speed and to prevent information loss that may occur for structures with a large number of hidden layers.
In comparison to a simple DNN with the same number of hidden layers, the training of the residual network converges to its best performance more quickly, by up to 50\%, for this analysis.
The residual network consists of several residual blocks as well as input and output layers.
In the residual block, the input data is branched off into two streams, where one stream is passed through two hidden layers while the other bypasses them.
The two streams are then added into one tensor and fed into the next residual block.
The network consists of 11 hidden layers in total, with five residual blocks.

\begin{figure}[h!]
    \centering
    \includegraphics[width=0.325\textwidth]{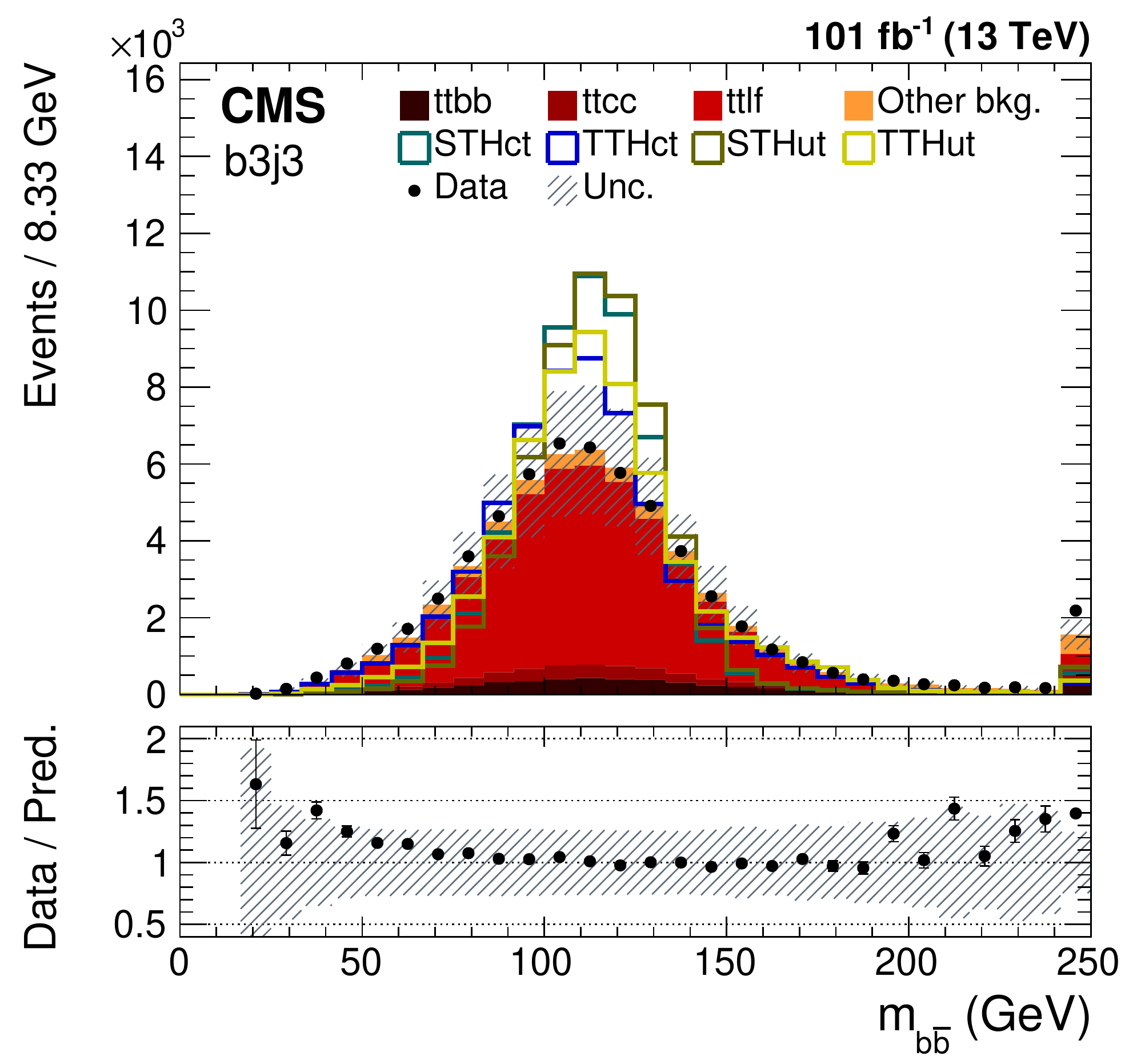}
    \includegraphics[width=0.325\textwidth]{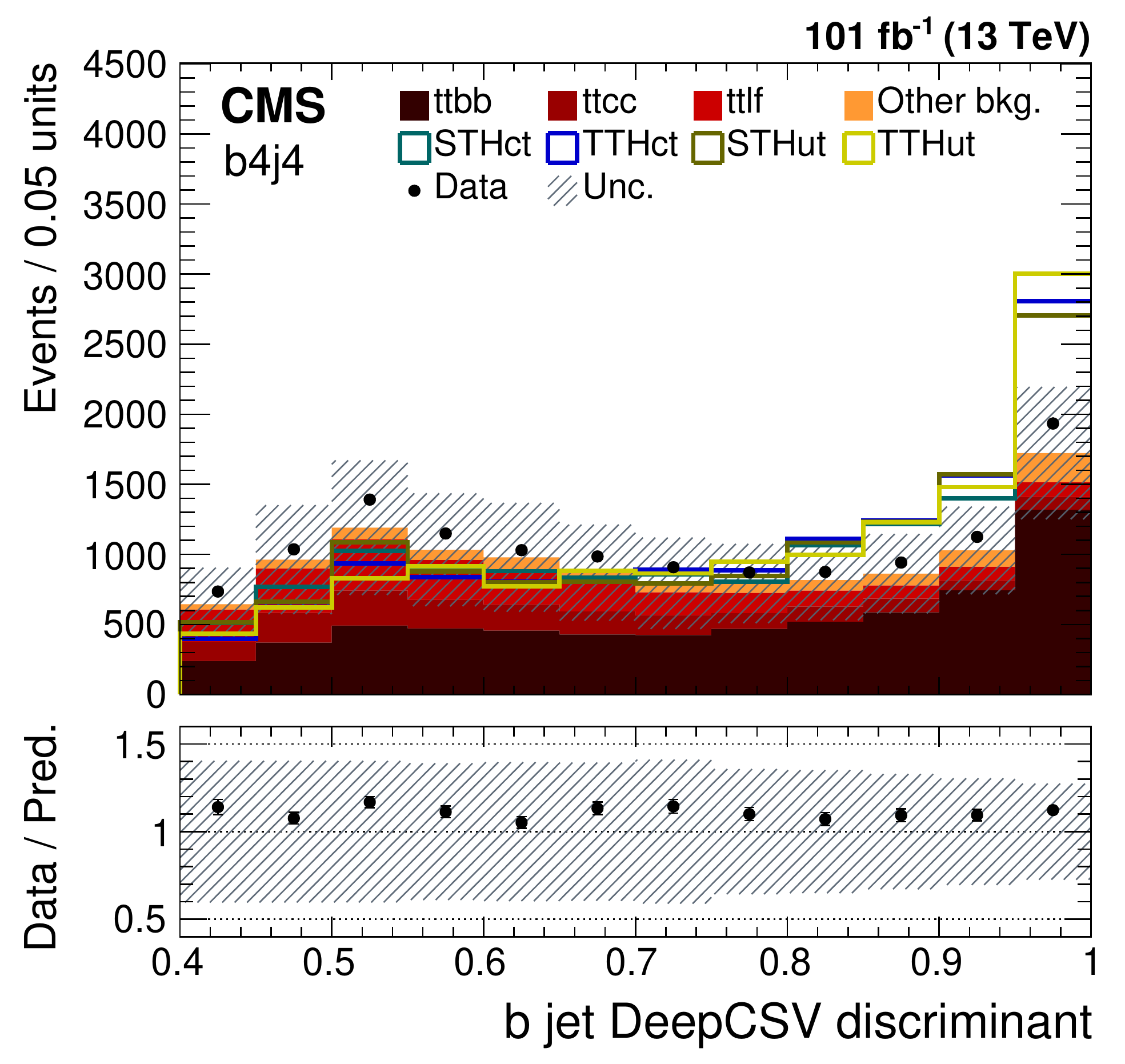}
    \includegraphics[width=0.325\textwidth]{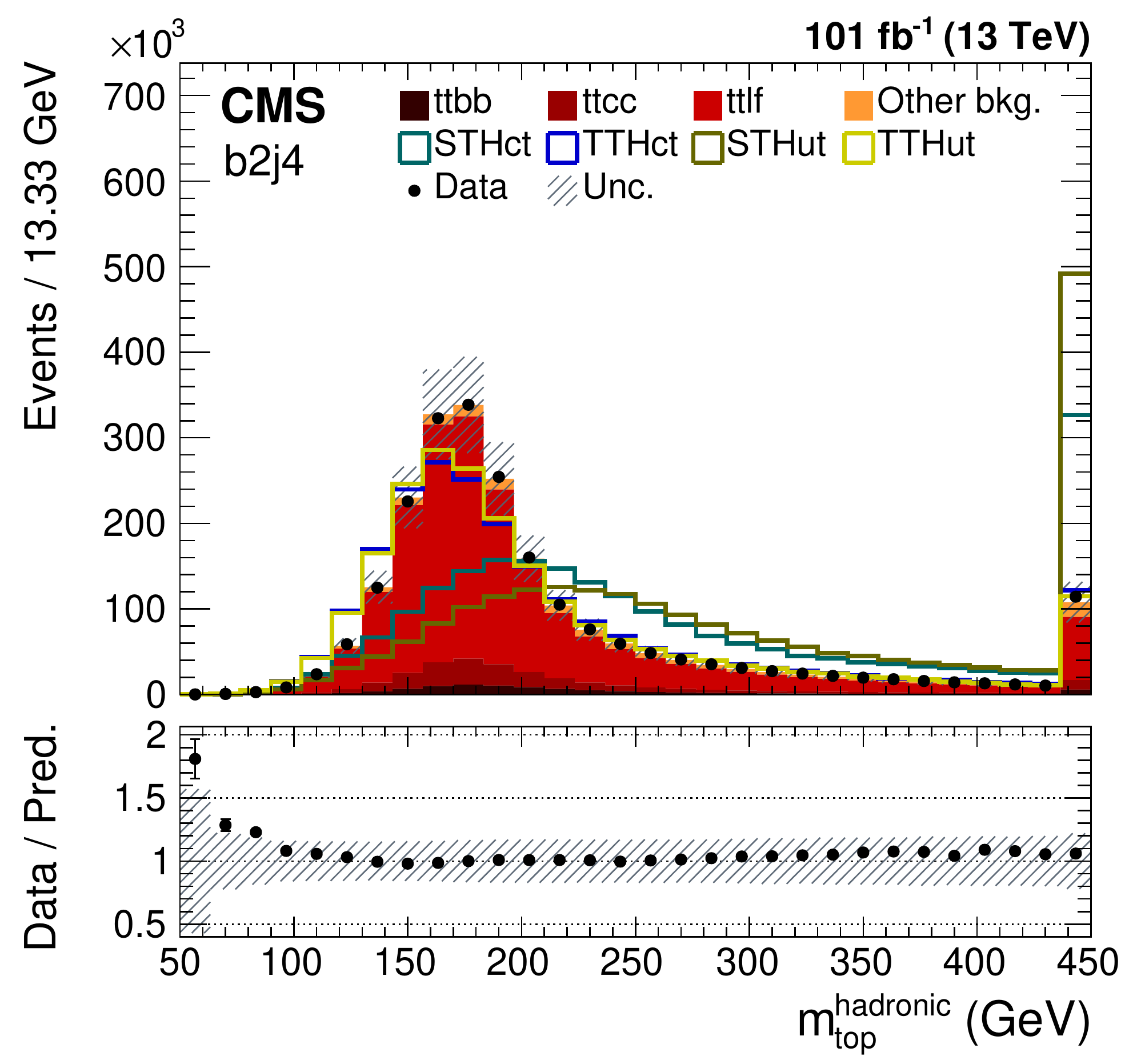}
    \caption{The reconstructed $m_{\bbbar}$ (Higgs boson candidate mass) in the b3j3 category for the ST signal scenario (left), the distribution of the second largest DeepCSV value for the \PQb-tagged jet from the Higgs boson decay in the b4j4 category for the TT signal scenario (middle), and the mass of the hadronically decaying top quark in the b2j4 category for the SM \ttbar background scenario (right) for the combined 2017+2018 data. The lower panel shows the ratio of observed data to the SM prediction. The shaded band corresponds to the total uncertainty in the predicted background. For the mass plots the last bin contains the overflow events.}
  \label{fig:hist_reco}
\end{figure}

The classifiers are first trained with all jet categories for the 2017 and 2018 simulated samples, separately.
In addition, an exclusive set of classifiers are trained using only the jet permutations from the b4j4 category because of the limited size for the signal combinations and the large difference between the signal and background sample sizes.
Once the classifiers are trained with the signal and background combinations, all combinations from the statistically independent data and simulated samples are predicted with the trained classifiers.
The combination with the highest DNN score is selected for each event and process hypothesis.
The performance of the classifier is measured in terms of the reconstruction efficiency, defined as the ratio of the number of events where all jets in the selected combination are correctly assigned to the number of events where the reconstructed jets are fully matched to generated jets within an angular distance of 0.4.
The computed efficiencies for the ST signal hypothesis are ${\approx}86\%$ and ${\approx}82\%$ for the \Hut and \Hct couplings, respectively.
In the TT signal scenario, the classifier trained with the \Hut (\Hct) coupling shows an ${\approx}79$ (${\approx}77\%$) efficiency, while for the SM \ttbar hypothesis it is ${\approx}78\%$.
By implementing the DNN, the reconstruction efficiencies are improved by 5--15\% compared to the kinematic fit method that was used in the 2016 data analysis~\cite{CMS_FCNC_1}.
While the efficiencies of the exclusive training in the b4j4 category are comparable to the inclusive ones for the ST and TT hypotheses, in this phase space the \ttbar reconstruction efficiency is improved by more than 40\%.
The efficiencies are computed with statistically independent samples with respect to the training ones.
The improvements in the reconstruction algorithms benefit the analysis by helping to accurately construct the kinematic distributions.
Example kinematic distributions for different process hypotheses are presented in Fig.~\ref{fig:hist_reco}.

\begin{figure}[htb!]
	\centering
	\includegraphics[width=0.45\textwidth]{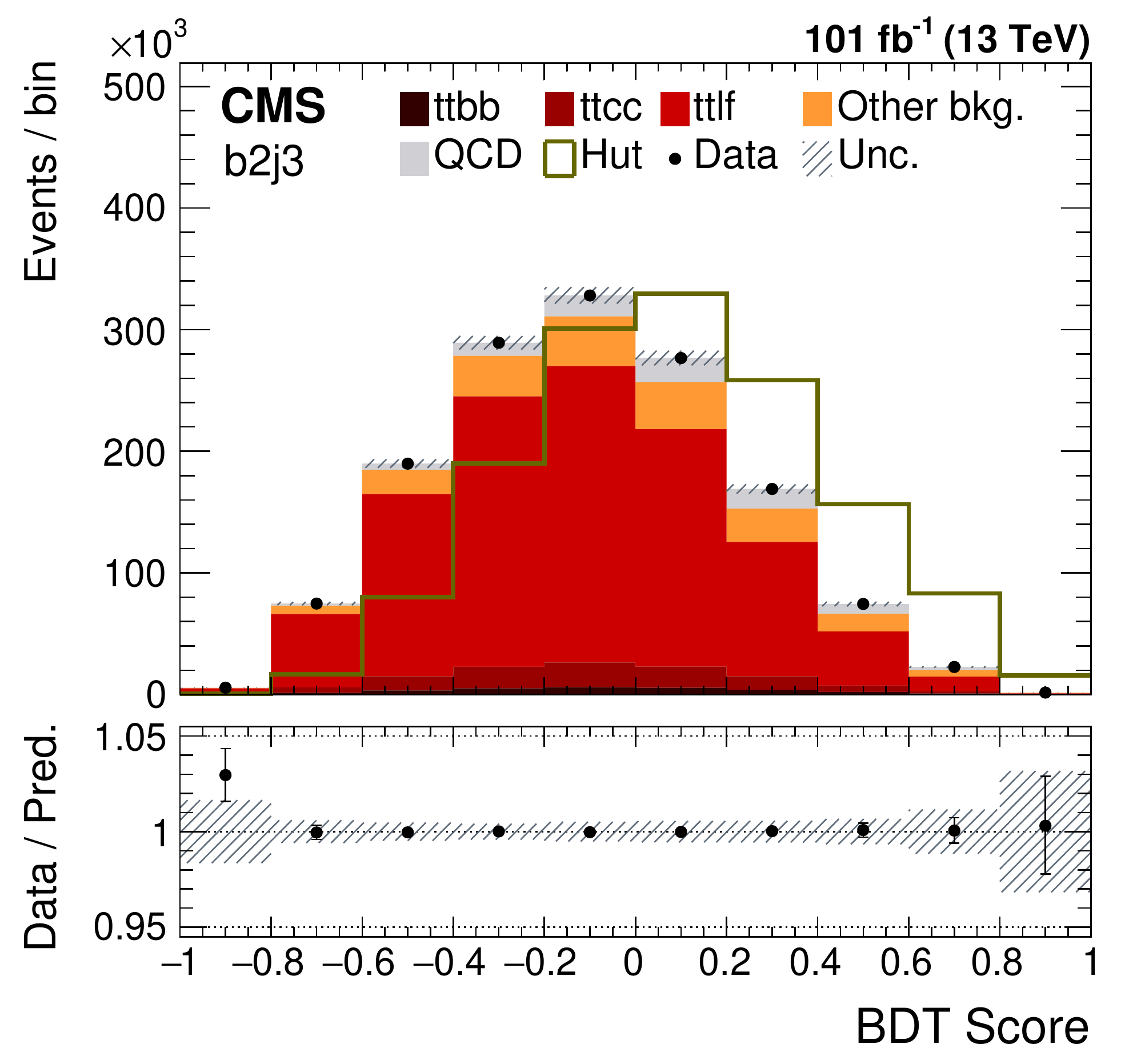}
	\includegraphics[width=0.45\textwidth]{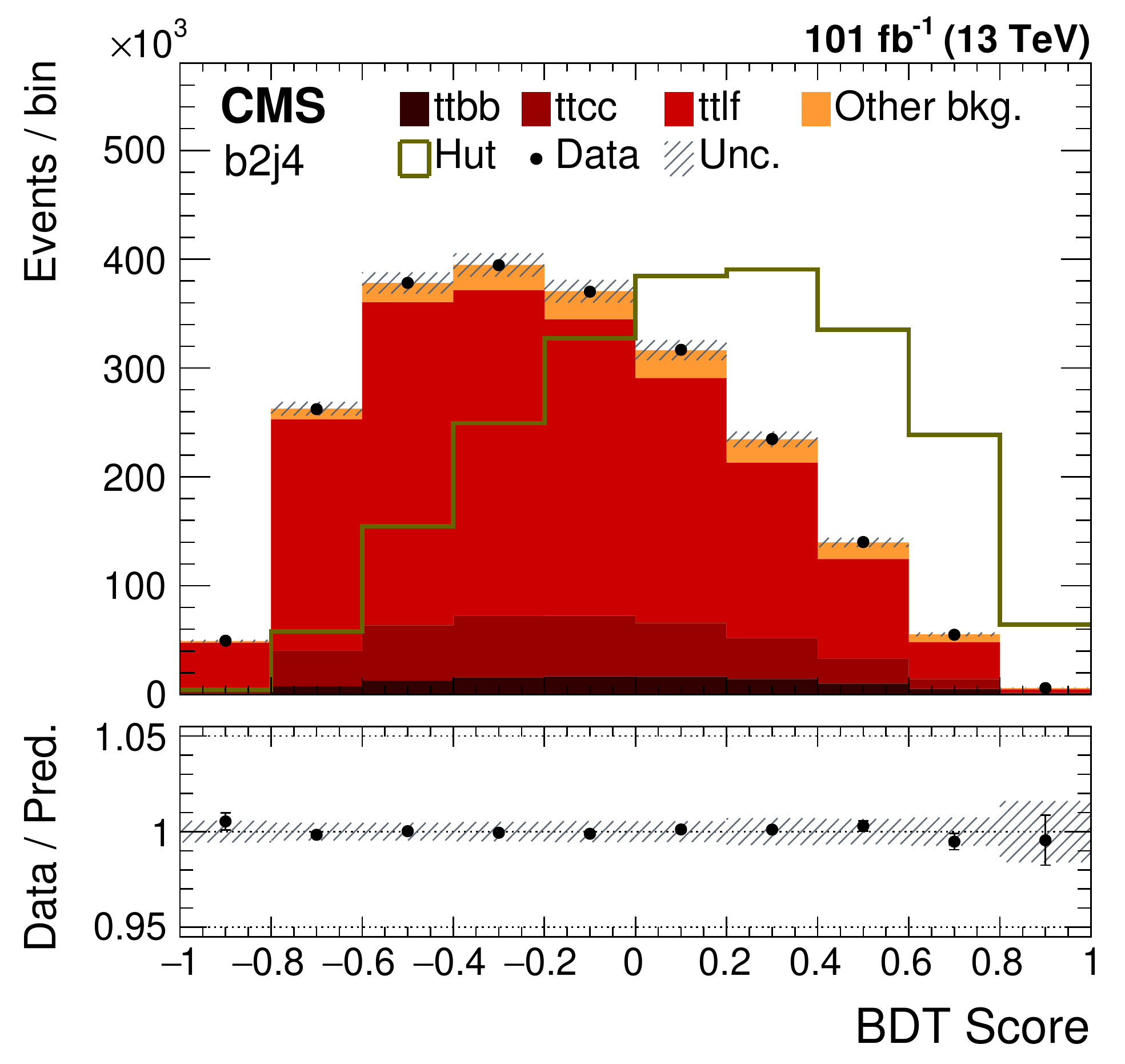}\\
	\includegraphics[width=0.45\textwidth]{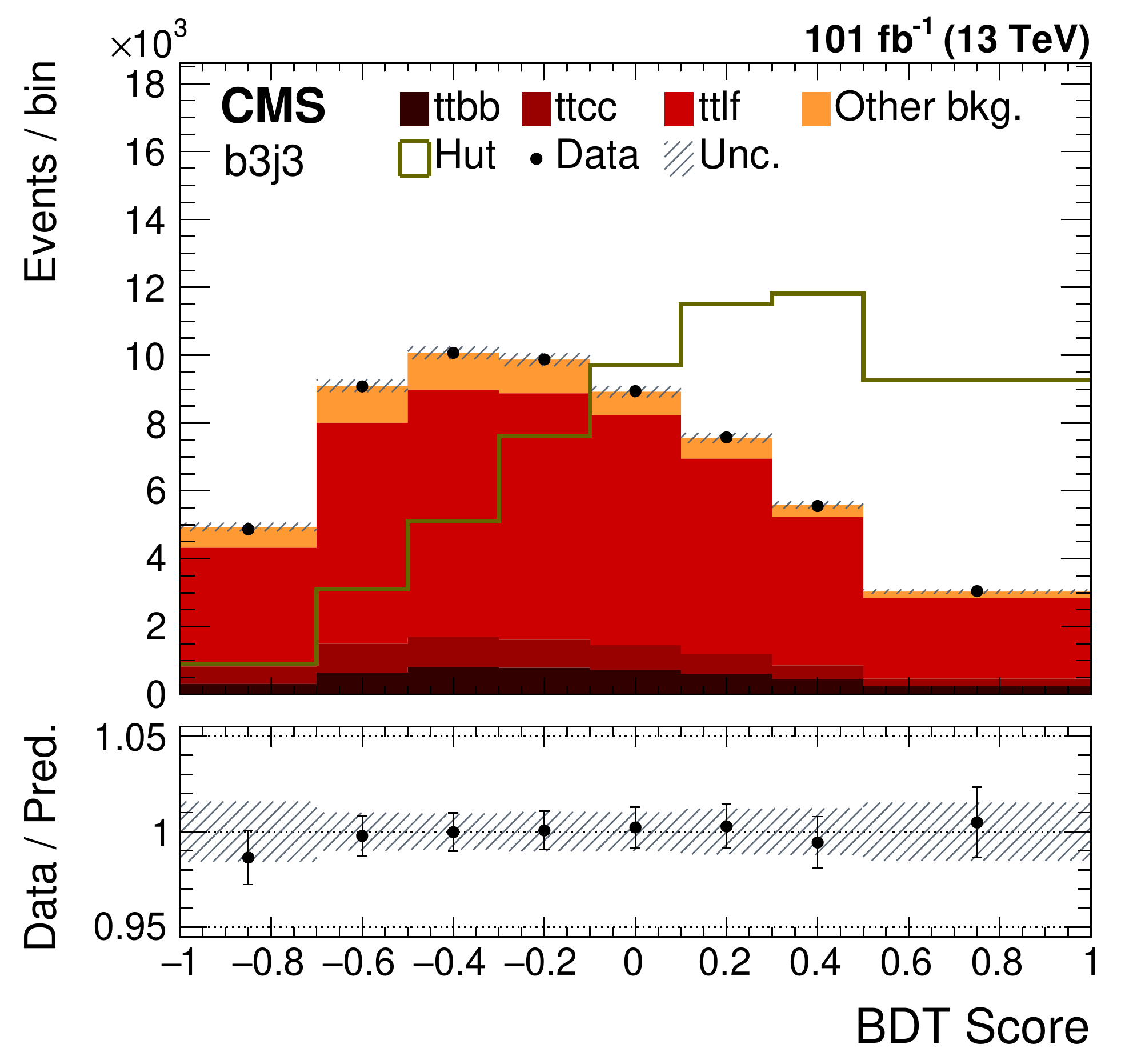}
	\includegraphics[width=0.45\textwidth]{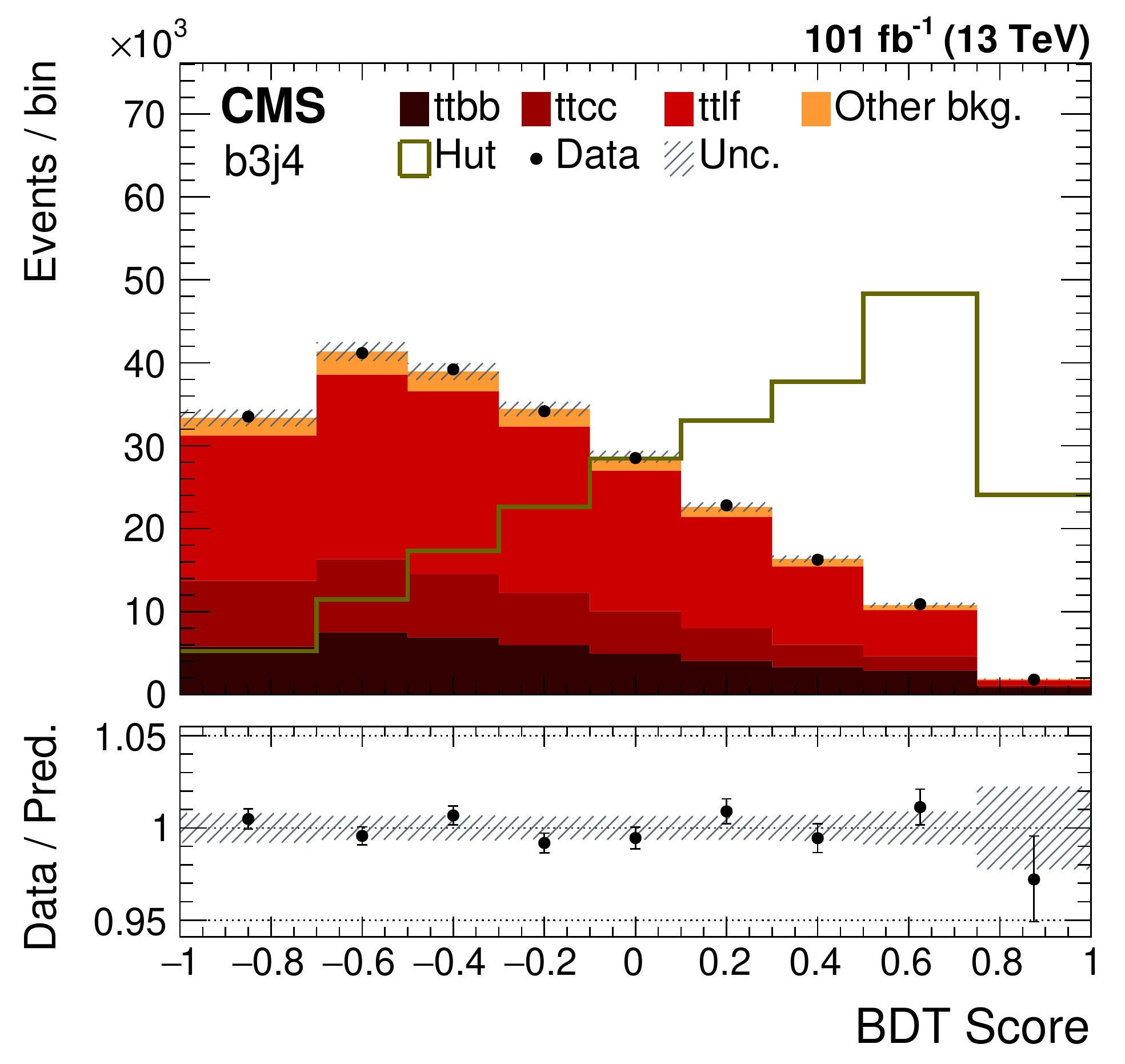}\\
	\includegraphics[width=0.45\textwidth]{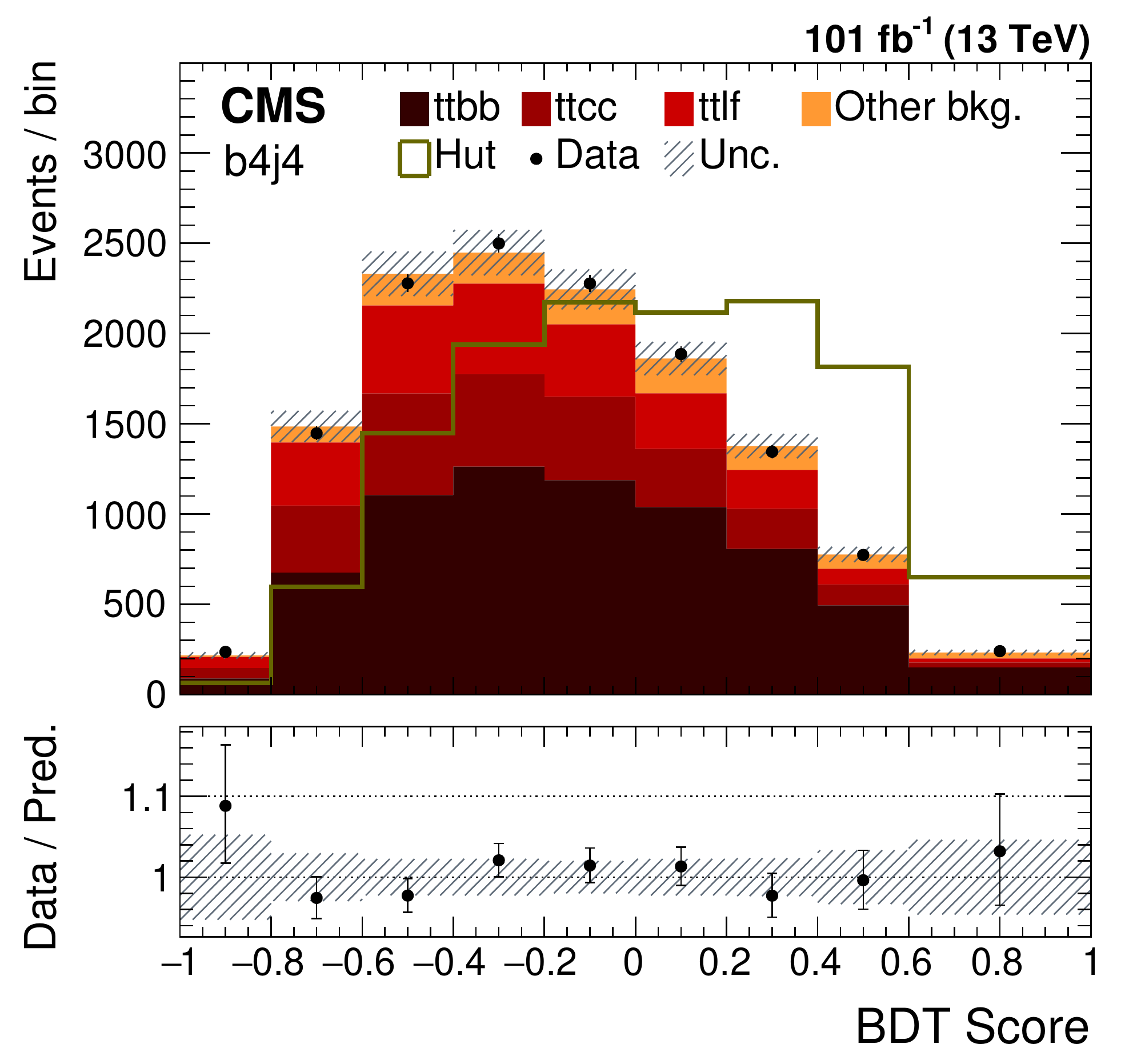}
	\caption{The BDT output distributions for the combined 2017+2018 data and simulation for the different jet categories, assuming the \Hut coupling. The lower panel shows the ratio of observed data to the SM prediction. The shaded bands correspond to the post-fit total uncertainty in the predicted background. The signal contributions are normalized to the total number of events in data.}
    \label{fig:postfitBDTHut}
\end{figure}

\begin{figure}[htb!]
	\centering
	\includegraphics[width=0.45\textwidth]{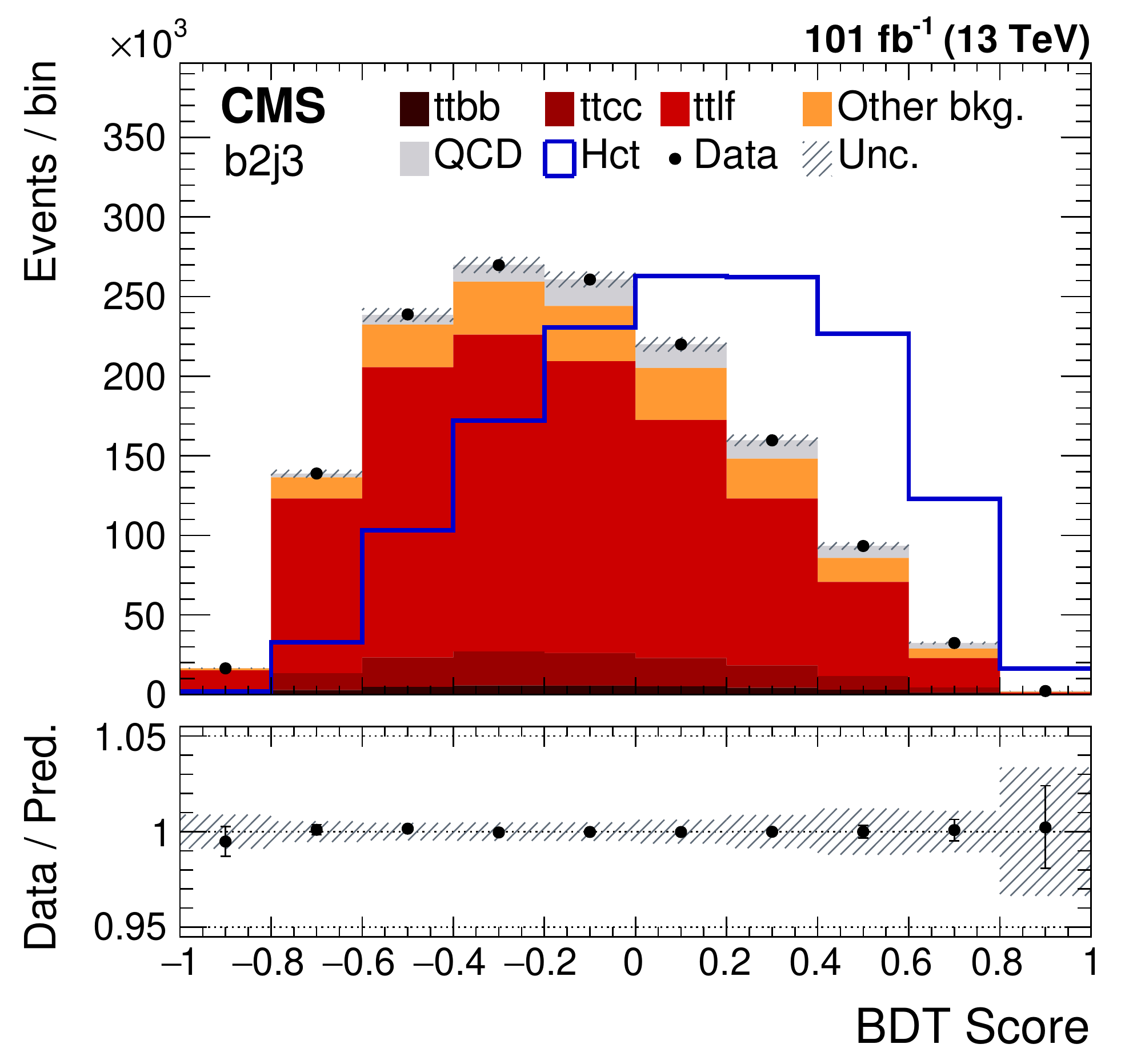}
	\includegraphics[width=0.45\textwidth]{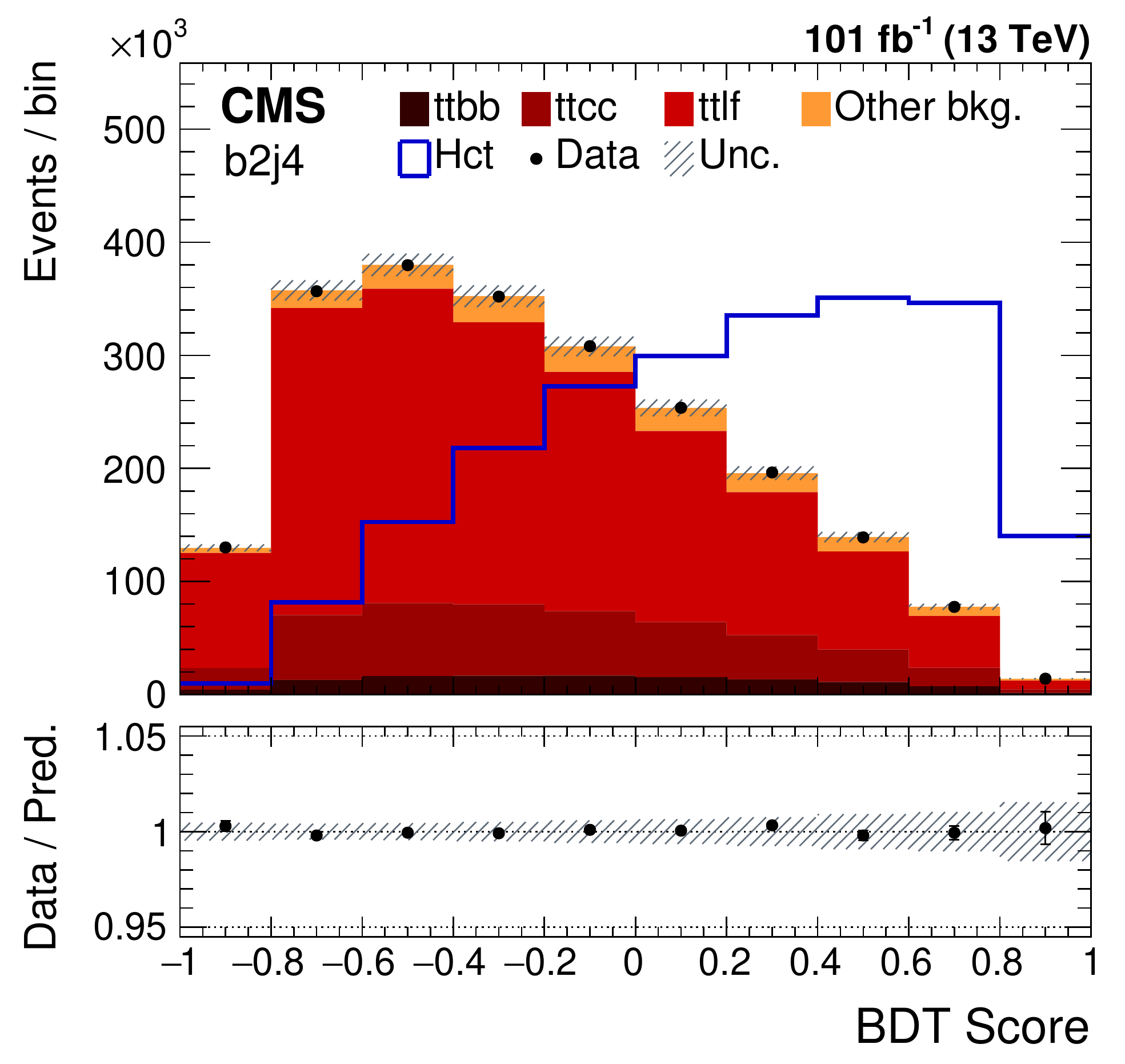}\\
	\includegraphics[width=0.45\textwidth]{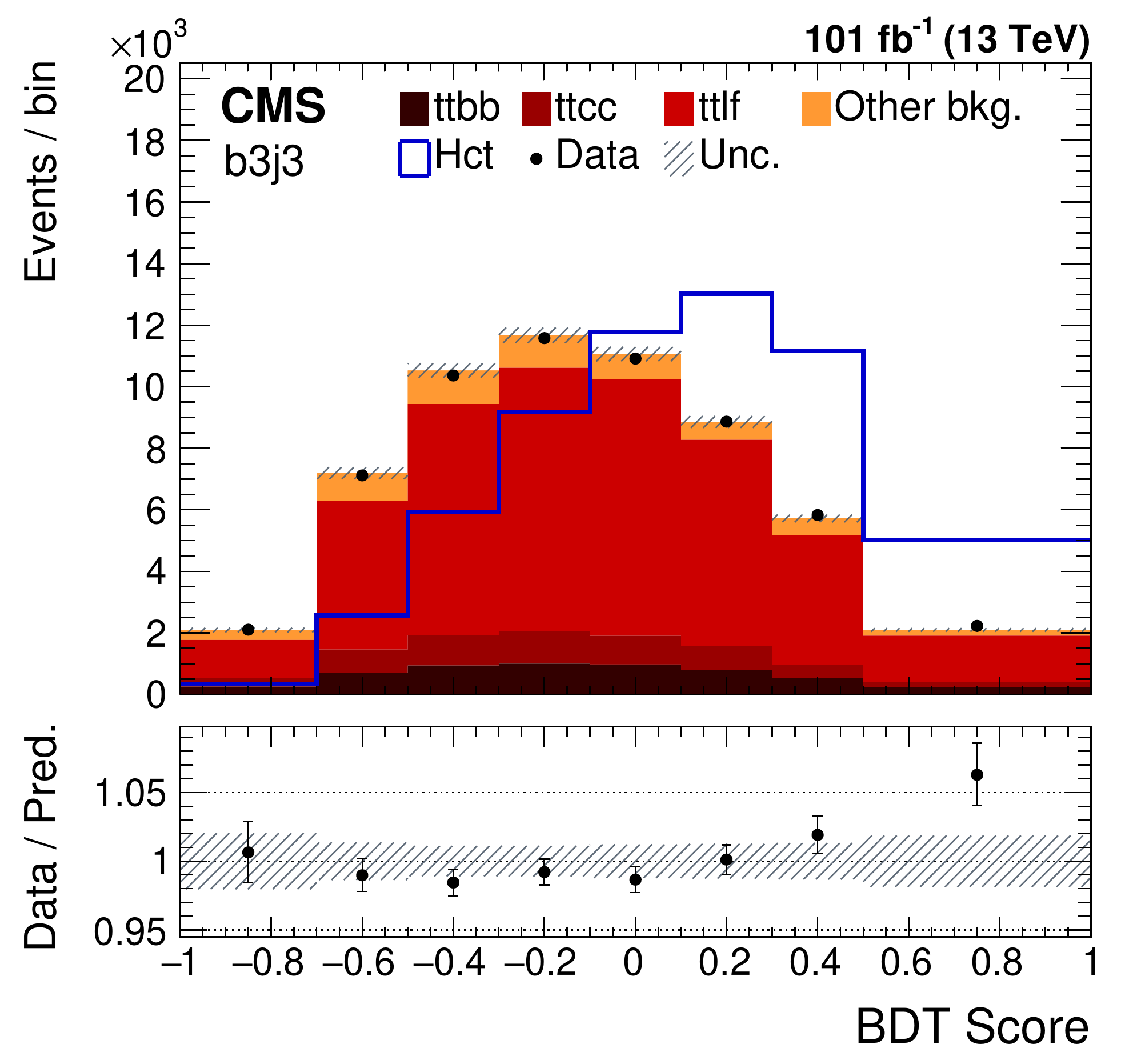}
	\includegraphics[width=0.45\textwidth]{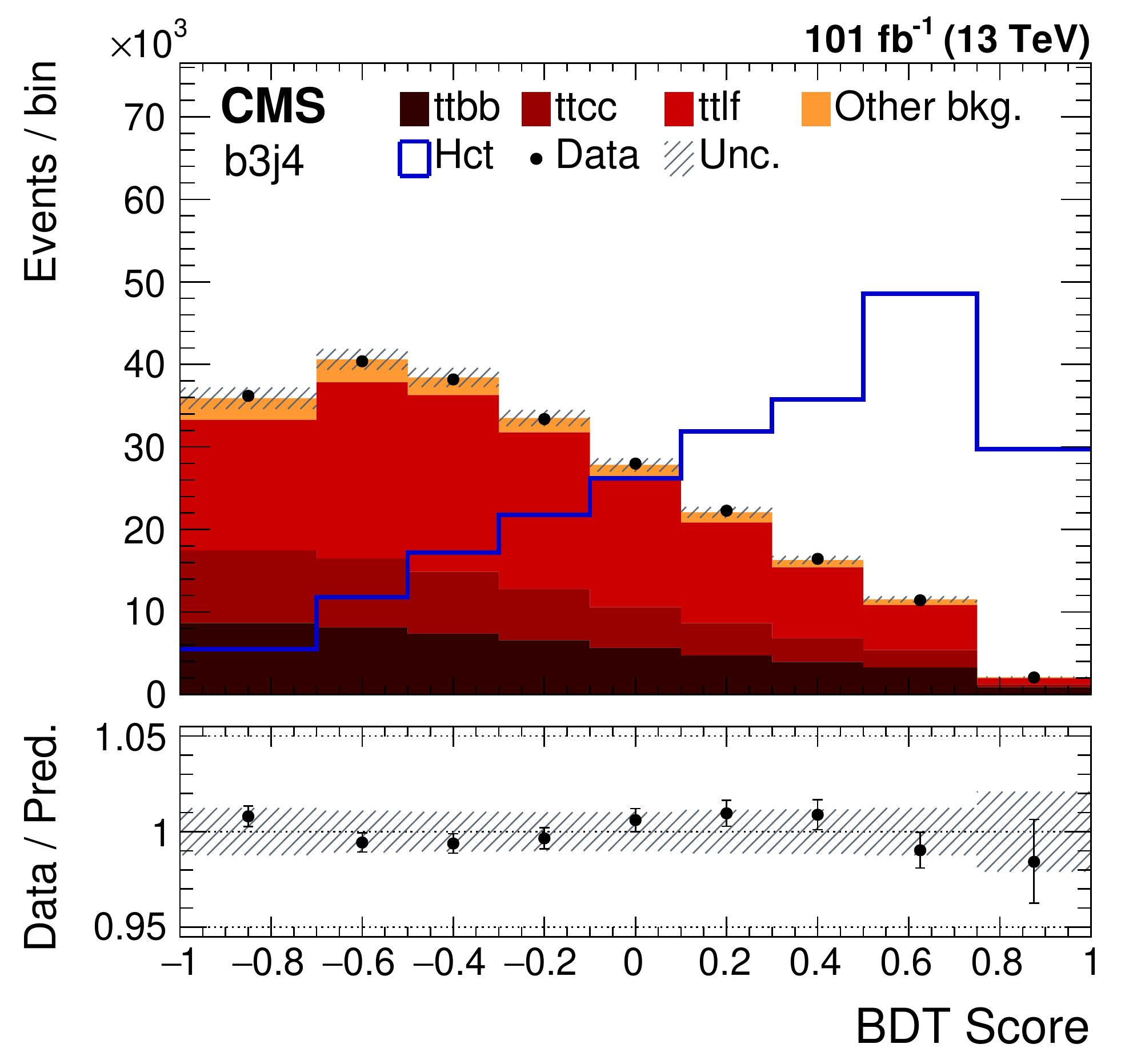}\\
	\includegraphics[width=0.45\textwidth]{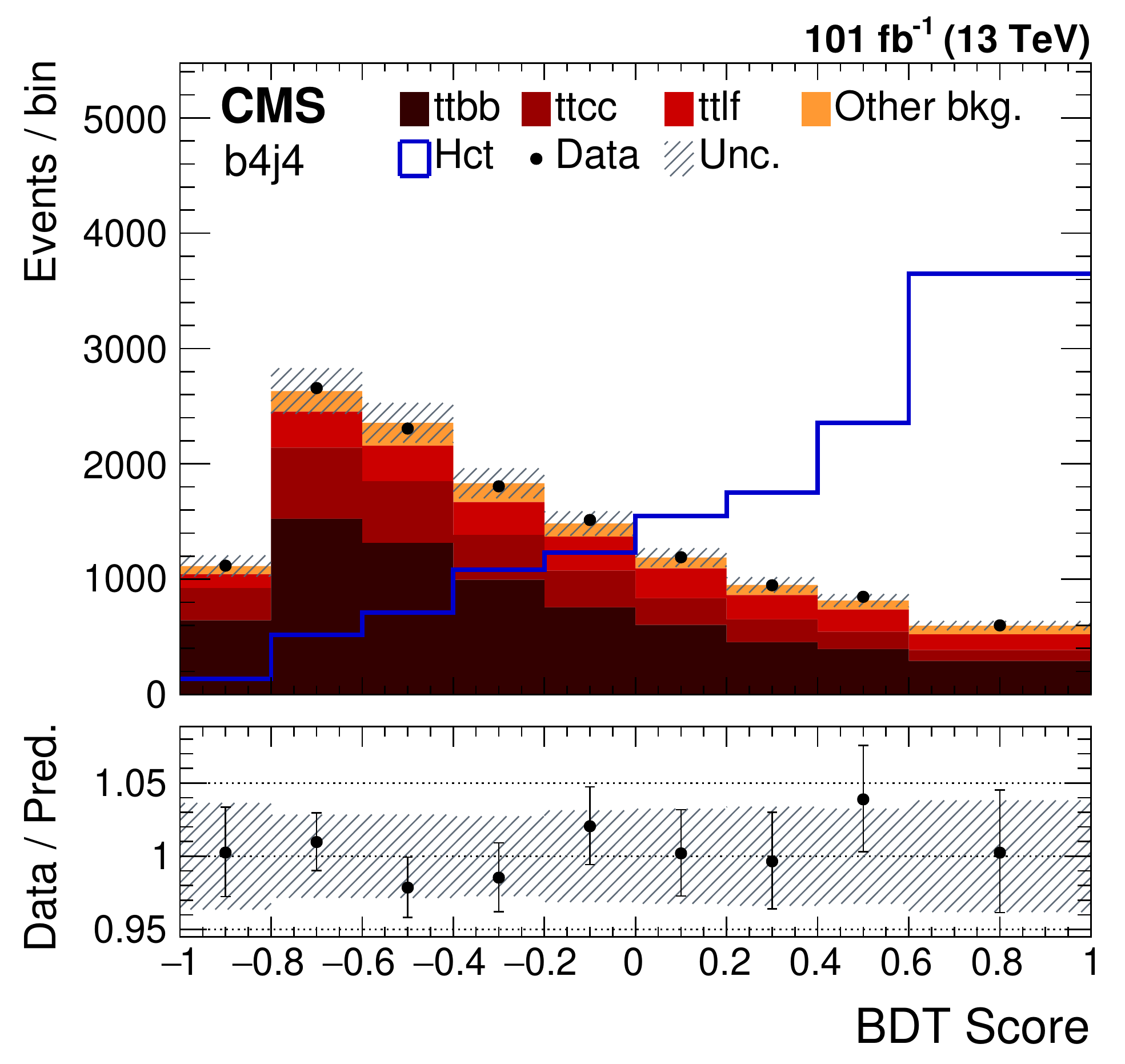}
	\caption{The BDT output distributions for the combined 2017+2018 data and simulation for the different jet categories, assuming the \Hct coupling. The lower panel shows the ratio of observed data to the SM prediction. The shaded bands correspond to the post-fit total uncertainty in the predicted background. The signal contributions are normalized to the total number of events in data.}
    \label{fig:postfitBDTHct}
\end{figure}

To distinguish signal events from the SM background contributions, a BDT is used for each signal coupling, jet category, and data-taking period (2017 and 2018).
The ST and TT FCNC events are considered jointly as signal in the training, taking into account their relative contributions in the event yield.
Among the background processes, only the dominant SM \ttbar process is considered in the BDT training.
A total of 164 input kinematic variables sensitive to the different process hypotheses, such as the \PQb tagging discriminator scores of the selected jets or the angular distributions between particles, are used in the BDT training.
Throughout the BDT training, \PQb tagging discriminator values are found to be the most important features.
For example, the distributions of the second largest DeepCSV values from the reconstructed Higgs boson in the ST and TT signal hypotheses are well separated from the SM predictions (see Fig.~\ref{fig:hist_reco}), as the corresponding jet typically does not originate from a bottom quark for the SM \ttbar hypothesis.
Meanwhile, once all the \PQb-tagged jets are assigned to the \PQb jets from the SM top quarks in the SM \ttbar hypothesis, the largest DeepCSV value from the reconstructed \PW boson is typically larger in ST and TT events compared to the SM events.
In addition, the mass distributions are also important features.
If there are more than three \PQb-tagged jets so that there are combinatorial jet permutations, the mass distribution of the reconstructed Higgs boson in FCNC events are distinguishable from the SM backgrounds.
Furthermore, the mass distribution of the hadronically decaying top quark in ST events is well separated from the background as the latter contains no such top quark.
In the b2j3 category, the BDT scores are additionally evaluated for the multijet process as they have nonnegligible contributions to the event yield.
The post-fit BDT output distributions for different jet categories for the 2017 and 2018 data-taking periods are obtained from the binned maximum likelihood fit to the data performed with the combined 2017 and 2018 data and simulation, and presented in Figs.~\ref{fig:postfitBDTHut} (\Hut) and~\ref{fig:postfitBDTHct} (\Hct coupling).

\section{Systematic uncertainties}
\label{sec:systematics}

The impact of experimental and theoretical systematic uncertainties is evaluated by means of nuisance parameters, described further in Section~\ref{sec:results}.
The dominant source of experimental uncertainty comes from the imperfect knowledge of the \PQb tagging (mis-)identification rate, which affects both the event yield in each analysis category and the discriminant distributions.
Its magnitude is evaluated separately for \PQb jets and light-quark or gluon jets, and is derived based on a \ttbar (\PZ) enriched sample for the former (latter).
The total magnitude of the \PQb tagging uncertainty varies with the jet category and ranges from 12 to 25\%.

The uncertainty arising from the jet energy scale (JES) and resolution (JER) also impacts both the distributions and the event yields for the signal and background processes.
The JES uncertainties are calculated by splitting into each one of the uncertainty sources, such as pileup, detector response, discrepancies in jet \pt and $\eta$, and flavor.
The uncertainties are then grouped into seven sources defined according to their correlations and detector geometry.
The uncertainty induced by the JER is assessed for different sources, such as the statistical uncertainty in the JER measurement, jet radiation, underlying event, out-of-cone showering, and pileup.
A detailed description of the determination of these uncertainties can be found in Ref.~\cite{CMS:2011shu}.
The uncertainties in the muon and electron scale factors related to their identification, isolation, and trigger efficiency are assessed when deriving these scale factors using the tag-and-probe technique, taking various systematic sources into account, including the limited size of the sample used, the possible bias from the tag selection, as well as run-by-run differences.
The experimental uncertainties in the integrated luminosity measurements for the data recorded in 2017 and 2018, respectively, are estimated to be 2.3 and 2.5\%~\cite{CMS-PAS-LUM-17-004,CMS-PAS-LUM-18-002}, while the integrated luminosity of the combined 2017--2018 data set has an uncertainty of 2.0\%.
To account for the uncertainty in the pileup reweighting in the simulated events, the value of the total $\Pp\Pp$ inelastic cross section, which is used to estimate the mean number of additional $\Pp\Pp$ interactions, is varied by $\pm$4.6\%~\cite{CMS:2018mlc}.

The following sources of theoretical uncertainty are taken into account.
The leading source of theoretical uncertainties is coming from the arbitrary choice of renormalization and factorization scales and is obtained by taking the envelope of the distributions obtained with modified scales~\cite{Cacciari:2003fi,Catani:2003zt}.
The scale variation uncertainty ranges between 10--17\% and constitutes the second largest uncertainty source.
The uncertainties in the initial- and final-state radiation are computed by varying the scale used to evaluate \alpS for each radiation by factors of 0.5 and 2.
The matching of matrix-element partons and parton showers is controlled in \POWHEG by a model parameter \hdamp.
The nominal value for the \hdamp parameter is set to 1.379\,\mtop, with $\mtop=172.5\GeV$~\cite{Sirunyan:2019dfx}.
The uncertainty associated with this prescription is estimated by varying the \hdamp value by 0.8738\,\mtop and 2.305\,\mtop.
The impact of the uncertainty in the PDF is assessed by using various event weights that represent the usage of the uncertainty eigenvector sets of the PDF~\cite{Butterworth:2015oua}.
Dedicated samples related to the variation of the underlying event tune CP5~\cite{Sirunyan:2019dfx} of the SM \ttbar background sample are used to estimate the associated uncertainty.

The calculation of the SM \ttbar production cross section has an uncertainty of 5.5\%~\cite{Nason:2004rx, Frixione:2007vw, Alioli:2010xd}.
An uncertainty of about 20\% in the cross section of the \ttbb process is considered to account for the correction of the underestimated simulation~\cite{Sirunyan:2020kga}, and an uncertainty of 50\% in the cross section of the \ttcc process is included to cover the possible discrepancy between data and simulation.
The uncertainties in the normalization of the other SM background yields have been found to be subdominant and are fixed to 10\%.
The uncertainty coming from the limited size of the simulated samples is also considered as a source of systematic uncertainty by assigning a single nuisance parameter that is constrained by the total uncertainty in each bin of the BDT output distribution~\cite{Barlow:1993dm,Conway:2011in}.

\section{Results}
\label{sec:results}

The BDT output distributions in Figs.~\ref{fig:postfitBDTHut} and \ref{fig:postfitBDTHct} show no significant excess with respect to the SM background expectations. Upper limits at 95\% confidence level (\CL) are placed on the product of the signal production cross section and branching fraction.
The limits are derived based on the asymptotic modified frequentist method (asymptotic \CLs) with the test statistic based on the profile likelihood ratio consists of binned likelihood as a function of signal strength and nuisance parameters~\cite{JUNK1999435, Read_2002, Cowan2011}, separately for the \ttoHu and \ttoHc couplings, \ie, assuming one nonvanishing coupling at a time.
All the systematic uncertainties described in Section~\ref{sec:systematics} are considered as nuisance parameters in the fit.
The uncertainty sources affecting the global normalization are considered with log-normal probability distributions, while the ones affecting the shape of the BDT distributions are considered with Gaussian probability distributions.
The BDT distributions from the five jet categories and from the three data-taking periods (the 2016 ones being taken from the analysis described in Ref.~\cite{CMS_FCNC_1}) are used to derive the final results.
Table \ref{tab:1718postfit} shows the number of events per process and jet category, derived from the simultaneous fit for the 2017+2018 data and simulation.

\begin{table}[ht]
   \centering
   \topcaption{Number of events in the combined 2017+2018 data and prediction for the backgrounds assuming absence of the signal events, shown separately for each jet category, with uncertainties obtained from the fit on the BDT distributions of \Hct coupling.}\label{tab:1718postfit}
   \setlength{\tabcolsep}{2pt}
   \cmsTable{
    \begin{tabular}{ccrclcrclcrclcrclcrcl}
      \hline
      Category  & & \multicolumn{3}{c}{b2j3} & & \multicolumn{3}{c}{b2j4} & & \multicolumn{3}{c}{b3j3} & & \multicolumn{3}{c}{b3j4} & & \multicolumn{3}{c}{b4j4} \\ \hline
      Data      & & \multicolumn{3}{c}{  1\,431\,931 } & & \multicolumn{3}{c}{  2\,207\,121 } & & \multicolumn{3}{c}{    59\,002 } & & \multicolumn{3}{c}{   228\,311 } & & \multicolumn{3}{c}{    12\,981 } \\ [1mm]
      \ttbb     & &    33\,038 & $\pm$ &  6500 & &   115\,300 & $\pm$ & 24\,000 & &     5452 & $\pm$ &   430 & &    49\,251 & $\pm$ &  3900 & &     6974 & $\pm$ &   710 \\
      \ttcc     & &   115\,968 & $\pm$ & 19\,000 & &   397\,302 & $\pm$ & 61\,000 & &     5362 & $\pm$ &   840 & &    45\,055 & $\pm$ &  6500 & &     2815 & $\pm$ &   830 \\
      \ttbar LF & &  1\,020\,955 & $\pm$ & 22\,000 & &  1\,547\,909 & $\pm$ & 47\,000 & &    42\,893 & $\pm$ &   990 & &   120\,172 & $\pm$ &  4500 & &     2118 & $\pm$ &   230 \\
      Other     & &   188\,582 & $\pm$ & 17\,000 & &   146\,711 & $\pm$ & 14\,000 & &     5508 & $\pm$ &   630 & &    13\,757 & $\pm$ &  1400 & &     1061 & $\pm$ &   120 \\
      QCD       & &    73\,167 & $\pm$ & 13\,000 & & & \NA & & & & \NA & & & & \NA & & & & \NA & \\ [1mm]
      Total     & &  1\,431\,710 & $\pm$ &  8900 & &  2\,207\,222 & $\pm$ & 13\,000 & &    59\,215 & $\pm$ &   750 & &   228\,235 & $\pm$ &  2500 & &    12\,967 & $\pm$ &   400 \\ \hline
   \end{tabular}
   }
\end{table}

\begin{figure}[htb!]
    \centering
    \includegraphics[width=0.49\textwidth]{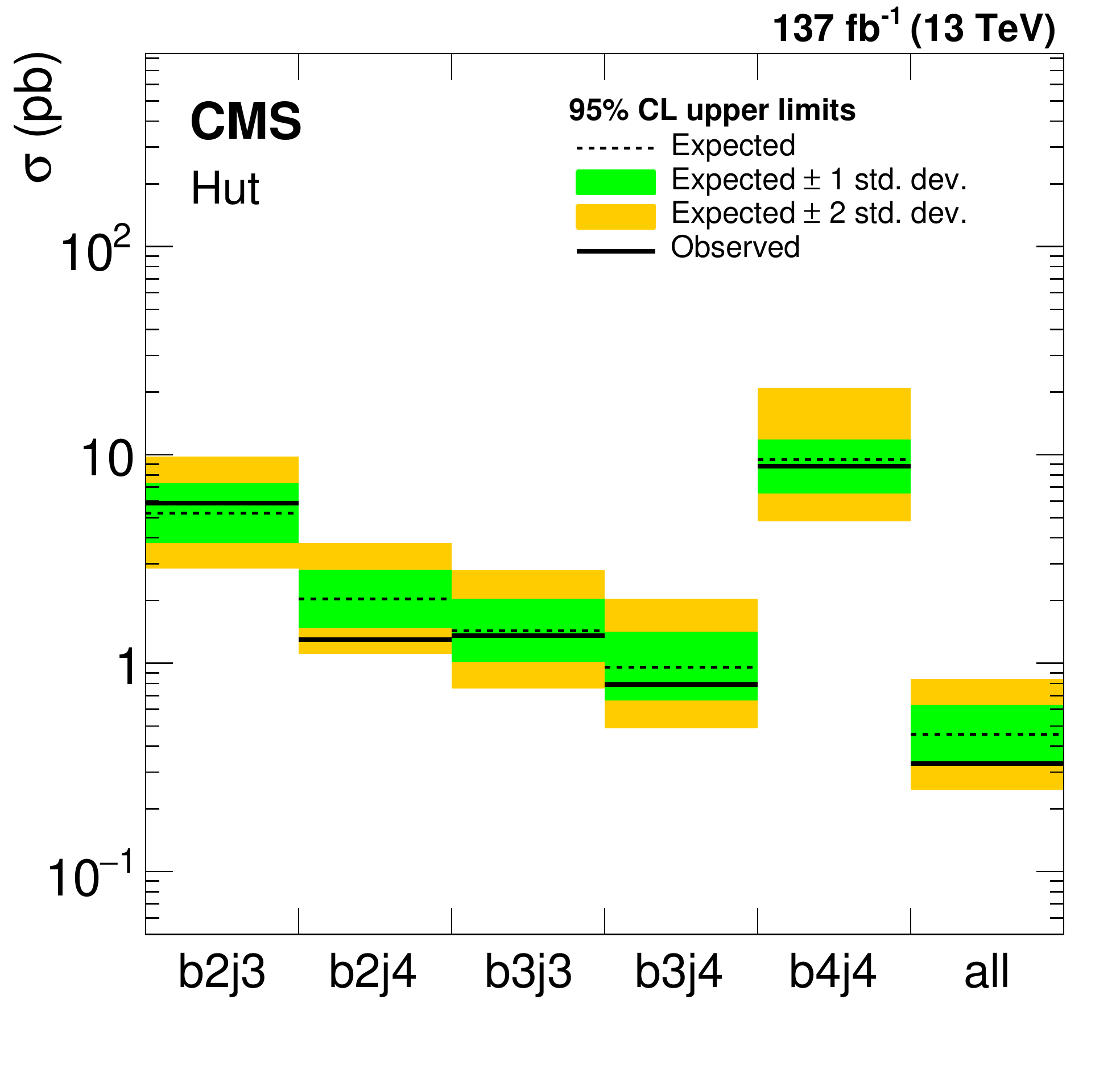}
    \includegraphics[width=0.49\textwidth]{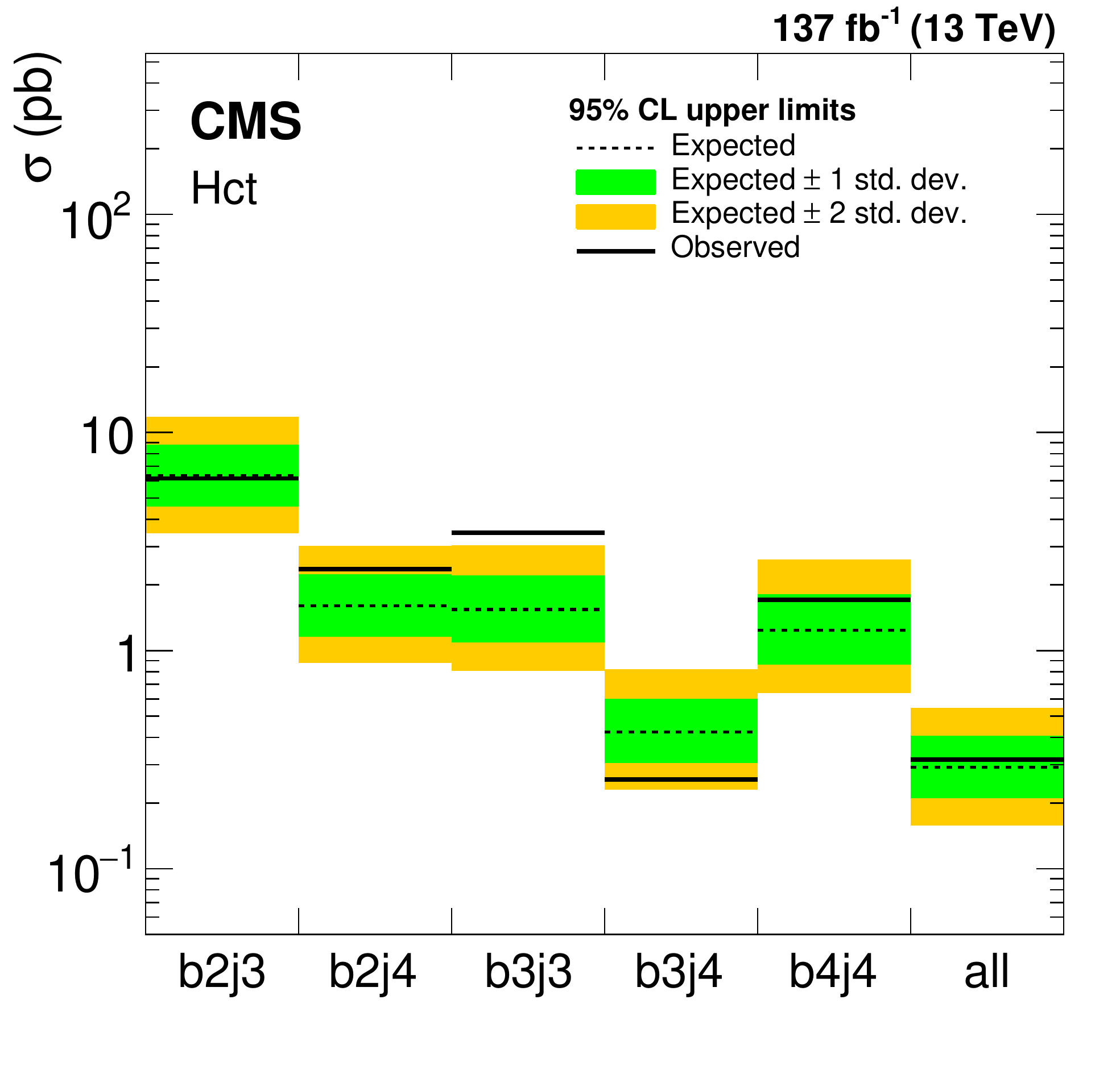}
    \caption{Excluded limits on the product of the cross section and branching fraction at 95\% \CL for the \Hut (left) and \Hct (right) couplings obtained using the BDT distributions. Each jet category and their combination are shown separately.}
  \label{fig:limits_allyear_jetsplitted}
\end{figure}

The expected and observed exclusion limits on the product of the cross section and branching fraction for the two couplings are shown in Fig.~\ref{fig:limits_allyear_jetsplitted} separately for each jet category as well as their combination.
The b3j4 category has the highest sensitivity to both couplings since it corresponds to the canonical jet content of the dominant TT signal contribution, while the SM \ttbar contribution is suppressed due to the requirement of a third \PQb jet.
The b2j3 and b4j4 categories show reversed sensitivity between \Hut and \Hct couplings due to the much higher \PQb jet misidentification probability of \PQc jets compared to \PQu jets.

\begin{figure}[htb!]
    \centering
    \includegraphics[width=0.49\textwidth]{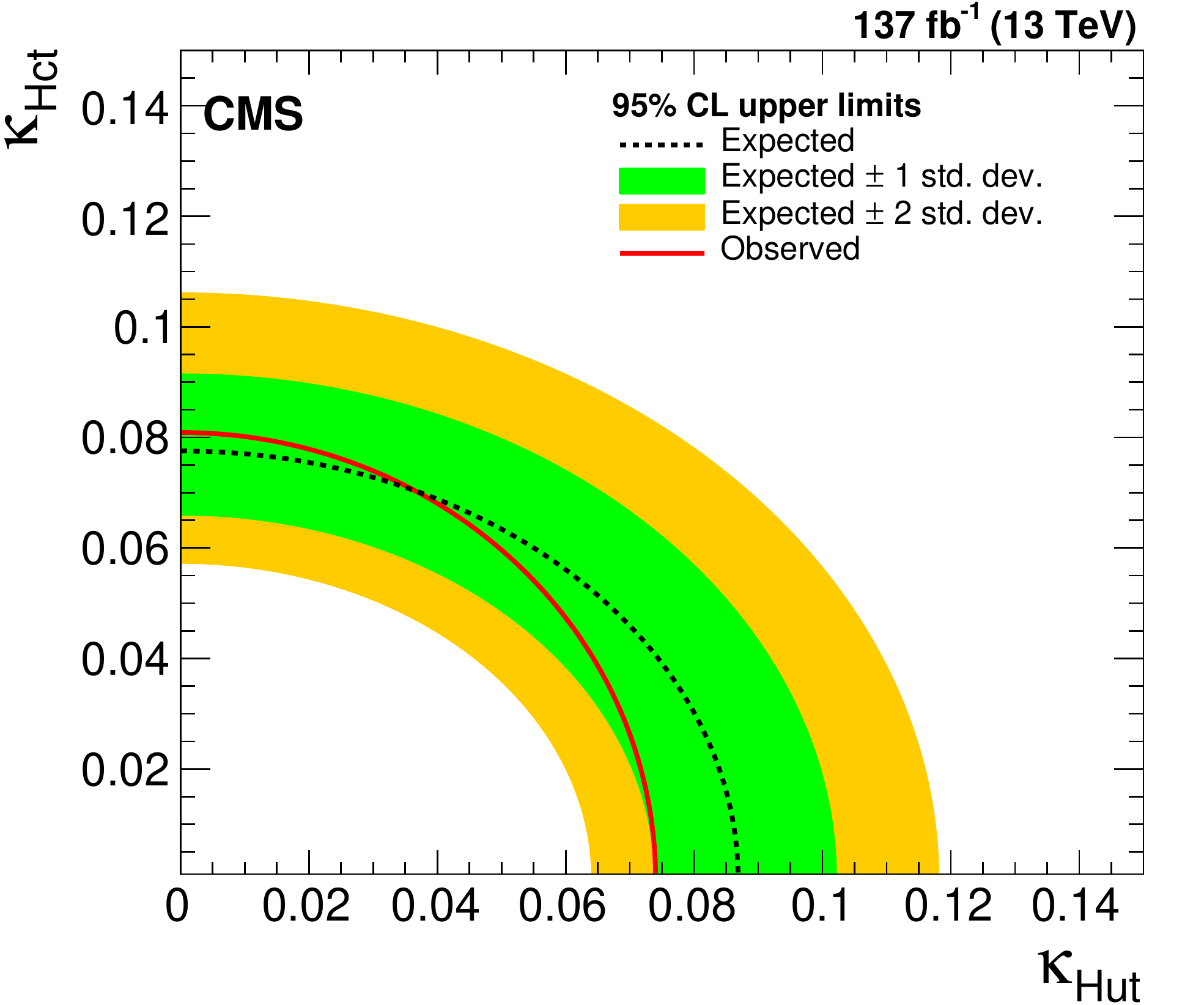}
    \includegraphics[width=0.49\textwidth]{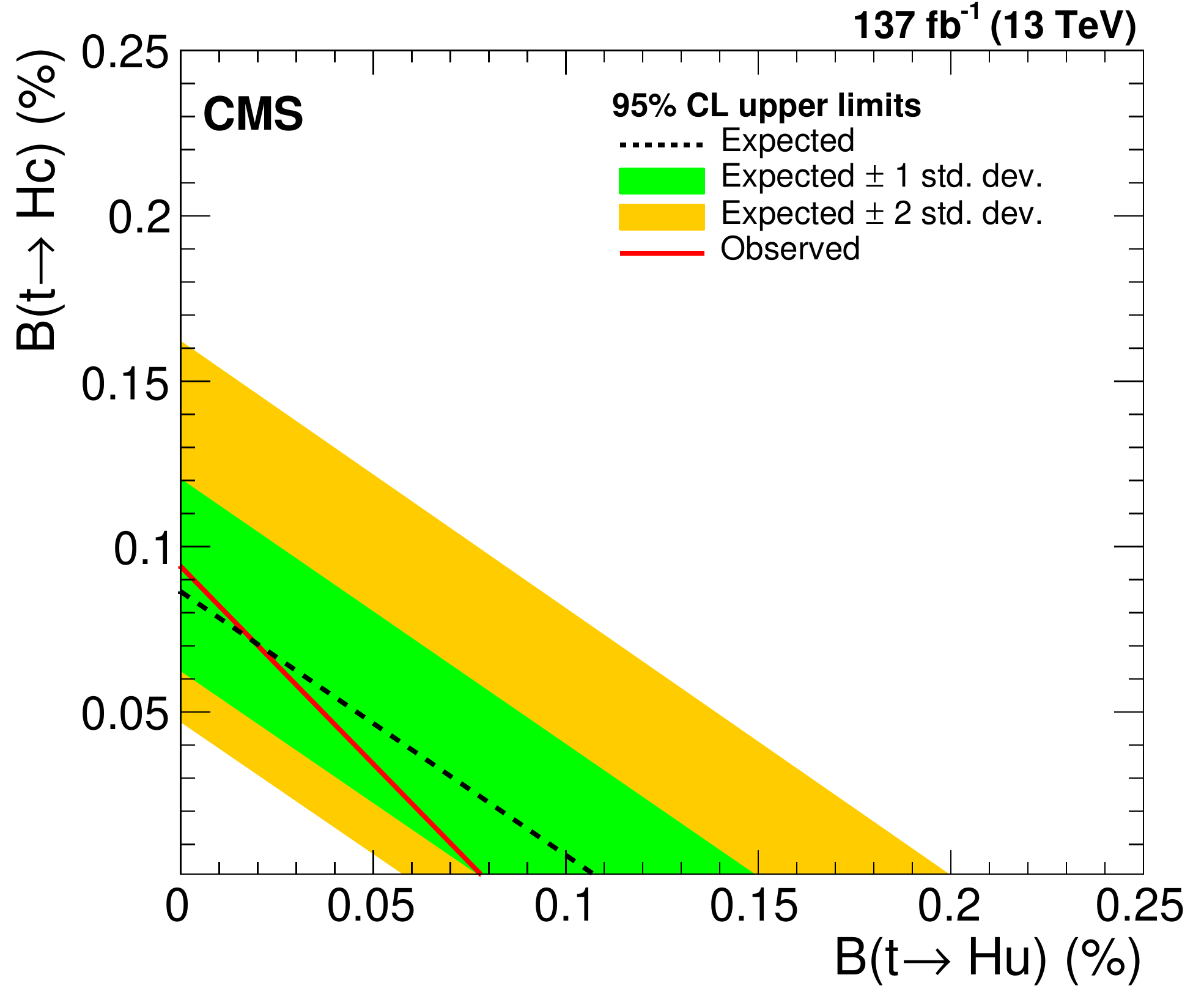}
    \caption{Upper limits on the couplings \kappaHut and \kappaHct (left), and the branching fractions $\mathcal{B}(\ttoHu)$ and $\mathcal{B}(\ttoHc)$ (right) at 95\% CL.}
  \label{fig:limits_coupling}
\end{figure}

The exclusion limit on the cross section can be interpreted in terms of an exclusion on the anomalous coupling \kappaHqt since the signal cross section scales linearly with $\kappaHqt^2$.
Using the signal cross sections for $\kappaHqt=1$ from Section~\ref{sec:samples} leads to observed (expected) limits of $\kappaHut < 0.074\,(0.087)$ and $\kappaHct < 0.081\,(0.078)$ at 95\% \CL.
This can in turn be interpreted in terms of a limit on the \ttoHq branching fraction thanks to the relation
\begin{equation}
  \kappaHqt^2 = \mathcal{B}(\ttoHq)\frac{\Gamma_{\PQt}}{\Gamma_{\Hqt}},
\end{equation}
with $\Gamma_{\PQt}=1.32\GeV$ at NNLO~\cite{topwidth} and $\Gamma_{\Hut}^{\kappaHut=1} = \Gamma_{\Hct}^{\kappaHct=1} = 0.19\GeV$. The observed (expected) 95\% \CL exclusion limits on the branching fractions are $\mathcal{B}(\ttoHu) < 0.079\,(0.11)\%$ and $\mathcal{B}(\ttoHc) < 0.094\,(0.086)\%$.
The limits for a single nonvanishing coupling are interpolated by assuming a linear relationship between the branching fractions.
The resulting two-dimensional limits for the branching fractions and couplings are shown in Fig.~\ref{fig:limits_coupling}.
The results presented here improve by a factor of 3--6 compared to those obtained in the same decay channel with the 2016 data set~\cite{CMS_FCNC_1}, and are comparable to the observed limits from the 13\TeV analysis in the diphoton channel~\cite{CMS:2021bdg}.

\section{Summary}
\label{sec:summary}

A search for flavor-changing neutral current interactions in events with a top quark (\PQt) decaying leptonically and a Higgs boson (\PH) decaying to a bottom quark-antiquark pair has been presented.
The search uses the LHC data, collected at $\sqrt{s} = 13\TeV$ in 2016--2018 and corresponding to an integrated luminosity of 137\fbinv of proton-proton collisions.
Events are analyzed, in the single-lepton channel containing a muon or electron in addition to the presence of at least three jets, where at least two of them are identified as originating from the hadronization of a bottom quark.
No significant deviation from the standard model prediction has been observed and upper limits on the branching fractions $\mathcal{B}(\ttoHq)$ have been set where \PQq refers to the up (\PQu) and charm quarks (\PQc);
their observed (expected) excluded values at 95\% confidence level are $\mathcal{B}(\ttoHu) < 0.079\,(0.11)\%$ and $\mathcal{B}(\ttoHc) < 0.094\,(0.086)\%$.
The observed limits reach the order of $10^{-4}$ in terms of branching fraction compared to the prediction from several well-known extensions of standard model that predict the values as high as $10^{-5}$ to $10^{-3}$.
This search substantially improves upon previous CMS results in the same final state by exploiting the larger integrated luminosity of 137\fbinv and by using advanced multivariate analysis techniques to perform the kinematic event reconstruction and signal extraction.

\begin{acknowledgments}
  We congratulate our colleagues in the CERN accelerator departments for the excellent performance of the LHC and thank the technical and administrative staffs at CERN and at other CMS institutes for their contributions to the success of the CMS effort. In addition, we gratefully acknowledge the computing centres and personnel of the Worldwide LHC Computing Grid and other centres for delivering so effectively the computing infrastructure essential to our analyses. Finally, we acknowledge the enduring support for the construction and operation of the LHC, the CMS detector, and the supporting computing infrastructure provided by the following funding agencies: BMBWF and FWF (Austria); FNRS and FWO (Belgium); CNPq, CAPES, FAPERJ, FAPERGS, and FAPESP (Brazil); MES and BNSF (Bulgaria); CERN; CAS, MoST, and NSFC (China); MINCIENCIAS (Colombia); MSES and CSF (Croatia); RIF (Cyprus); SENESCYT (Ecuador); MoER, ERC PUT and ERDF (Estonia); Academy of Finland, MEC, and HIP (Finland); CEA and CNRS/IN2P3 (France); BMBF, DFG, and HGF (Germany); GSRI (Greece); NKFIA (Hungary); DAE and DST (India); IPM (Iran); SFI (Ireland); INFN (Italy); MSIP and NRF (Republic of Korea); MES (Latvia); LAS (Lithuania); MOE and UM (Malaysia); BUAP, CINVESTAV, CONACYT, LNS, SEP, and UASLP-FAI (Mexico); MOS (Montenegro); MBIE (New Zealand); PAEC (Pakistan); MSHE and NSC (Poland); FCT (Portugal); JINR (Dubna); MON, RosAtom, RAS, RFBR, and NRC KI (Russia); MESTD (Serbia); MCIN/AEI and PCTI (Spain); MOSTR (Sri Lanka); Swiss Funding Agencies (Switzerland); MST (Taipei); ThEPCenter, IPST, STAR, and NSTDA (Thailand); TUBITAK and TAEK (Turkey); NASU (Ukraine); STFC (United Kingdom); DOE and NSF (USA).
   
  \hyphenation{Rachada-pisek} Individuals have received support from the Marie-Curie program and the European Research Council and Horizon 2020 Grant, contract Nos.\ 675440, 724704, 752730, 758316, 765710, 824093, 884104, and COST Action CA16108 (European Union); the Leventis Foundation; the Alfred P.\ Sloan Foundation; the Alexander von Humboldt Foundation; the Belgian Federal Science Policy Office; the Fonds pour la Formation \`a la Recherche dans l'Industrie et dans l'Agriculture (FRIA-Belgium); the Agentschap voor Innovatie door Wetenschap en Technologie (IWT-Belgium); the F.R.S.-FNRS and FWO (Belgium) under the ``Excellence of Science -- EOS" -- be.h project n.\ 30820817; the Beijing Municipal Science \& Technology Commission, No. Z191100007219010; the Ministry of Education, Youth and Sports (MEYS) of the Czech Republic; the Deutsche Forschungsgemeinschaft (DFG), under Germany's Excellence Strategy -- EXC 2121 ``Quantum Universe" -- 390833306, and under project number 400140256 - GRK2497; the Lend\"ulet (``Momentum") Program and the J\'anos Bolyai Research Scholarship of the Hungarian Academy of Sciences, the New National Excellence Program \'UNKP, the NKFIA research grants 123842, 123959, 124845, 124850, 125105, 128713, 128786, and 129058 (Hungary); the Council of Science and Industrial Research, India; the Latvian Council of Science; the Ministry of Science and Higher Education and the National Science Center, contracts Opus 2014/15/B/ST2/03998 and 2015/19/B/ST2/02861 (Poland); the Funda\c{c}\~ao para a Ci\^encia e a Tecnologia, grant CEECIND/01334/2018 (Portugal); the National Priorities Research Program by Qatar National Research Fund; the Ministry of Science and Higher Education, projects no. 14.W03.31.0026 and no. FSWW-2020-0008, and the Russian Foundation for Basic Research, project No.19-42-703014 (Russia); MCIN/AEI/10.13039/501100011033, ERDF ``a way of making Europe", and the Programa Estatal de Fomento de la Investigaci{\'o}n Cient{\'i}fica y T{\'e}cnica de Excelencia Mar\'{\i}a de Maeztu, grant MDM-2017-0765 and Programa Severo Ochoa del Principado de Asturias (Spain); the Stavros Niarchos Foundation (Greece); the Rachadapisek Sompot Fund for Postdoctoral Fellowship, Chulalongkorn University and the Chulalongkorn Academic into Its 2nd Century Project Advancement Project (Thailand); the Kavli Foundation; the Nvidia Corporation; the SuperMicro Corporation; the Welch Foundation, contract C-1845; and the Weston Havens Foundation (USA). 
\end{acknowledgments}

\bibliography{auto_generated}\cleardoublepage \appendix\section{The CMS Collaboration \label{app:collab}}\begin{sloppypar}\hyphenpenalty=5000\widowpenalty=500\clubpenalty=5000\cmsinstitute{Yerevan~Physics~Institute, Yerevan, Armenia}
A.~Tumasyan
\cmsinstitute{Institut~f\"{u}r~Hochenergiephysik, Vienna, Austria}
W.~Adam\cmsorcid{0000-0001-9099-4341}, J.W.~Andrejkovic, T.~Bergauer\cmsorcid{0000-0002-5786-0293}, S.~Chatterjee\cmsorcid{0000-0003-2660-0349}, K.~Damanakis, M.~Dragicevic\cmsorcid{0000-0003-1967-6783}, A.~Escalante~Del~Valle\cmsorcid{0000-0002-9702-6359}, R.~Fr\"{u}hwirth\cmsAuthorMark{1}, M.~Jeitler\cmsAuthorMark{1}\cmsorcid{0000-0002-5141-9560}, N.~Krammer, L.~Lechner\cmsorcid{0000-0002-3065-1141}, D.~Liko, I.~Mikulec, P.~Paulitsch, F.M.~Pitters, J.~Schieck\cmsAuthorMark{1}\cmsorcid{0000-0002-1058-8093}, R.~Sch\"{o}fbeck\cmsorcid{0000-0002-2332-8784}, D.~Schwarz, S.~Templ\cmsorcid{0000-0003-3137-5692}, W.~Waltenberger\cmsorcid{0000-0002-6215-7228}, C.-E.~Wulz\cmsAuthorMark{1}\cmsorcid{0000-0001-9226-5812}
% ...
\vskip\cmsinstskip
% ...
1:~Also at TU Wien, Wien, Austria\\
% ...
\end{sloppypar}
\end{document}